\documentclass[aps,prb,reprint,superscriptaddress]{revtex4-2}

\usepackage[dvipdfmx]{graphicx}
\usepackage[T1]{fontenc}
\usepackage{textcomp}
\usepackage{lmodern}
\usepackage{amsmath,amssymb,mathtools}
\usepackage{bm}
\usepackage{graphicx}
\usepackage{hyperref}
\usepackage{microtype}
\usepackage{siunitx}
\usepackage{physics} 
\usepackage{color}
\usepackage{comment}

\hypersetup{
	colorlinks=true,
	linkcolor=blue,
	citecolor=blue,
	urlcolor=blue, 
	breaklinks=true
}

\begin{document}

\title{Spin-orbital exchange as a route to intertwined dipole-quadrupole orbital order in MnV$_2$O$_4$ under strong trigonal crystal field}

\author{Hiroki Nakai}
\email{hiroki.nakai@utoronto.ca}
\affiliation{Graduate School of Arts and Sciences, University of Tokyo, Meguro-ku, Tokyo 153-8902, Japan}
\affiliation{Department of Physics, University of Toronto, 60 St. George Street, Toronto, Ontario, M5S 1A7, Canada}

\author{Yusuke Nomura}
\affiliation{Institute for Materials Research (IMR), Tohoku University, Sendai, 980-8577, Japan}
\affiliation{International Quantum Materials Center (Q$^2$), Advanced Institute for Materials Research (WPI-AIMR), Tohoku University, Sendai 980-8577, Japan}

\date{\today}

\begin{abstract}
Orbitally degenerate systems provide a promising platform for realizing novel quantum phases driven by spin-orbital exchange interactions, as described by the Kugel-Khomskii model. 
Spinel vanadates, in which orbital degrees of freedom remain active, exhibit structural and magnetic transitions accompanied by orbital ordering, but the nature of the orbital state in MnV$_2$O$_4$ remains under debate. 
Here, we combine first-principles calculations with an effective spin-orbital model to address this problem. 
We show that a significant trigonal crystal field is present in high-temperature cubic phase and plays an essential role in determining the low-energy degrees of freedom. 
Based on the resulting parameters, we construct an effective Hamiltonian beyond the conventional dominant-hopping approximation and demonstrate that subdominant hopping processes strongly modify the spin-orbital exchange interactions. 
As a result, the system stabilizes a two-in/two-out magnetic configuration featuring spin canting and intertwined dipole-quadrupole orbital order.
\end{abstract}

\maketitle

\section*{Introduction}

Orbital degrees of freedom provide an important source of diversity in the phases and physical properties of transition-metal compounds~\cite{Khomskii_book_2014}. 
In Mott insulators with high crystal symmetry, where orbital degeneracy is preserved, orbital fluctuations can couple spin and orbital degrees of freedom and give rise to intertwined orders. 
Such orbital fluctuations in these systems are described by the Kugel-Khomskii model~\cite{Kugel1973, Kugel1975}, in which spin-orbital exchange interactions emerge from virtual hopping processes. 
These interactions determine orbital configurations and, in turn, strongly influence the magnetic properties of the system~\cite{Kugel1982, Tokura2000, Khaliullin2005, Streltsov2017, Khomskii2022}.

The form of spin-orbital exchange interactions in the Kugel-Khomskii framework is strongly constrained by the anisotropic nature of the underlying orbital wave functions and the local crystal geometry, leading to bond-dependent and anisotropic couplings~\cite{Khaliullin2005}. 
Such interactions can stabilize a variety of ordered phases~\cite{Kugel1982, Brink2001, Mostovoy2002, Mochizuki2004, Jackeli2008, Jackeli2009_Sr2VO4} as well as quantum disordered phases~\cite{Feiner1997, Kugel2015, Koch-Janusz2015, Savary2021, Churchill2025}. 
This mechanism further underlies the emergence of highly anisotropic exchange interactions in strongly spin-orbit-coupled Mott insulators, as exemplified by the Kitaev model~\cite{Jackeli2009}. 
From this perspective, understanding how local crystal environments control the form and hierarchy of anisotropic spin-orbital exchange interactions is essential for identifying and designing novel quantum phases.

While exchange interactions play a central role, the spin-orbit coupling (SOC) and Jahn-Teller effect can further compete, giving rise to rich and complex physical behavior in $t_{2g}$ orbital systems~\cite{Khaliullin2005}.  
Spinel vanadates $A$V$_2$O$_4$ provide a prototypical platform for exploring such physics~\cite{Radaelli2005, Lee2010, Takagi2011, Tsurkan2021}. 
In these compounds, V$^{3+}$ ions host two electrons in the $t_{2g}$ orbitals; the orbital degeneracy is preserved in the high-temperature cubic phase, giving rise to active orbital degrees of freedom in addition to spin. 
Upon cooling, successive structural and magnetic transitions are observed at comparable temperatures, accompanied by the development of orbital order~\cite{Lee2004_ZnV2O4, Wheeler2010_MgV2O4, Katsufuji2008_FeV2O4, Suzuki2007_MnV2O4, Garlea2008_MnV2O4}. 
In contrast to $e_g$ orbital systems, where structural and magnetic transitions typically occur at well-separated temperature scales, this behavior highlights the role of competing microscopic energy scales in stabilizing ordered states.

From a theoretical perspective, the nature of the orbital order has been discussed in terms of either real-orbital~\cite{Tsunetsugu2003, Motome2004, Motome2005} or complex-orbital configurations~\cite{Tchernyshyov2004, Matteo2005}, depending on the competition among exchange interactions, SOC, and the Jahn-Teller effect.
These two scenarios predict different space-group symmetries, namely $I4_1/a$ and $I4_1/amd$, respectively. 
Despite these distinct theoretical predictions, a consensus on the nature of the orbital order has not yet been reached. 
In particular, in MnV$_2$O$_4$, X-ray diffraction studies have reported both $I4_1/amd$~\cite{Adachi2005_MnV2O4} and $I4_1/a$~\cite{Nii2012_MnV2O4}, leaving the issue unresolved.

For MnV$_2$O$_4$, first-principles calculations have suggested the presence of a significant trigonal crystal field and supported the $I4_1/a$ space group~\cite{Sarkar2009}. 
A subsequent model study incorporating this effect has also reproduced the same symmetry~\cite{Chern2010}. 
In this model, the exchange interactions are constructed from dominant hopping processes based on the spatial anisotropy of the $t_{2g}$ orbitals, as in earlier theoretical approaches~\cite{Tsunetsugu2003}. 
However, trigonal crystal fields reorganize the orbital states, suggesting that the resulting exchange processes can be more complex than those captured in such simplified descriptions.
Further support for the relevance of trigonal crystal fields in the cubic phase has been obtained from valence electron density analyses based on synchrotron X-ray diffraction, first in the related compound FeV$_2$O$_4$ ~\cite{Manjo2022, Koyama_2026} and, more recently, in MnV$_2$O$_4$ itself~\cite{Koyama2026}.

Motivated by these observations, we revisit this problem from the perspective of a microscopic model and investigate the orbital order and magnetic structure of MnV$_2$O$_4$. 
Since the orbital state is determined by competing energy scales, a quantitative evaluation of the material parameters is essential. 
We therefore perform first-principles calculations for the high-temperature cubic phase to determine these parameters, including the trigonal crystal field. 
Building on this, we construct an effective spin-orbital Hamiltonian that goes beyond the conventional dominant-hopping approximation and captures the full structure of the exchange interactions.
We show that subdominant hopping processes, which have been neglected in previous studies, significantly modify the exchange interactions and play a key role in determining the orbital order and magnetic structure. 
Moreover, starting from the cubic phase, our approach naturally accounts for the instability toward the tetragonal phase without invoking additional assumptions. 
Our results identify a spin-orbital state consistent with the $I4_1/amd$ space group and provide a unified microscopic understanding of the exchange mechanisms underlying the ordering phenomena in MnV$_2$O$_4$. 
This highlights the importance of local crystal environments in shaping exchange interactions and, in turn, stabilizing diverse spin-orbital phases in correlated materials.

\begin{figure}[t]
\centering
\includegraphics[width=7cm]{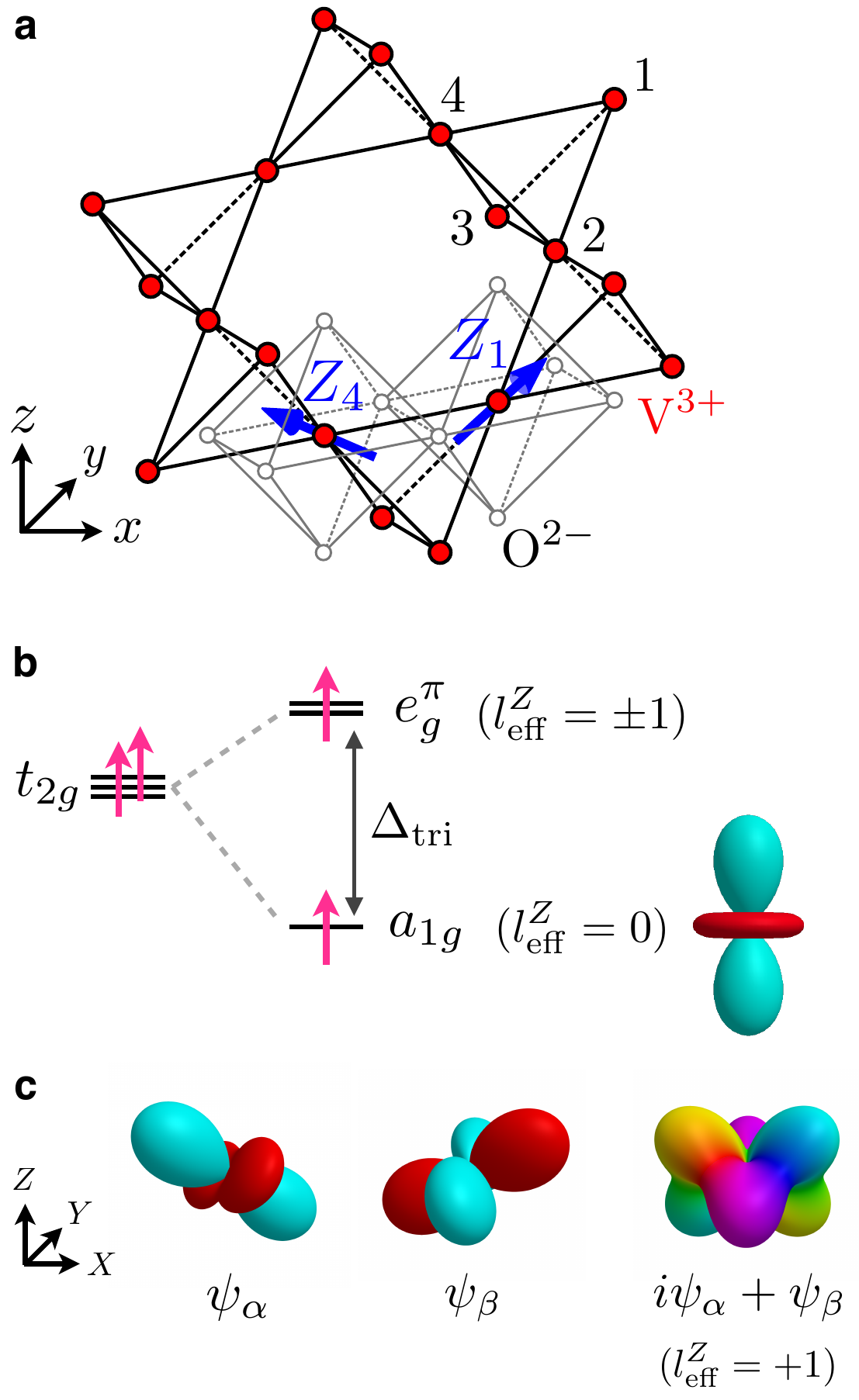}
\caption{
\textbf{Crystal structure and low-energy degrees of freedom in MnV$_2$O$_4$.}
\textbf{a} Spinel structure showing the pyrochlore lattice formed by V ions and edge-sharing VO$_6$ octahedra. 
The local trigonal axes are indicated by blue arrows. 
\textbf{b} Splitting of the $t_{2g}$ orbitals under the trigonal crystal field into an $a_{1g}$ singlet and an $e_g^\pi$ doublet, along with a schematic representation of the $d^2$ electron configuration at the V site. 
The twofold orbital degree of freedom in the $d^2$ manifold is described by an orbital pseudospin-1/2 $\tau$ (see text for details).
The $a_{1g}$ orbital wave function is also illustrated, with positive and negative isosurfaces depicted in turquoise-blue and red, respectively.
\textbf{c} One-electron orbital wave functions in the $e_g^\pi$ manifold, showing the real orbitals $\psi_\alpha$ and $\psi_\beta$ and the complex $l_{\rm eff}^Z=+1$ state formed by their linear combination.
The color scale represents the complex phase of the wave function.
}
\label{fig:structure}
\end{figure}

\section*{Results}

\subsection*{From multiorbital Hamiltonian to effective spin-orbital model}

We construct an effective spin-orbital model for the cubic phase starting from a multiorbital Hamiltonian. 
We first derive the relevant parameters from first-principles calculations and clarify the hierarchy of the associated energy scales. 
Based on this hierarchy, we next derive a Kugel-Khomskii-type effective Hamiltonian, which serves as the basis for the analysis presented below.

\subsubsection*{Multiorbital Hamiltonian}

To describe the low-energy physics of MnV$_2$O$_4$, we consider the multiorbital Hamiltonian for the $t_{2g}$ manifold,
\begin{equation}
\mathcal{H} = \mathcal{H}_{\rm C} + \mathcal{H}_{\rm CEF}^{\rm tri} + \mathcal{H}_{\rm SOC} + \mathcal{H}_{\rm kin},
\end{equation}
where $\mathcal{H}_{\rm C}$ denotes the on-site Coulomb interaction, $\mathcal{H}_{\rm CEF}^{\rm tri}$ the trigonal crystal field, $\mathcal{H}_{\rm SOC}$ the atomic SOC, and $\mathcal{H}_{\rm kin}$ the electron hopping between neighboring V ions. 
The explicit forms of these terms are provided in Sec.~S1 of the Supplementary Materials (SM).
The two-body interaction parameters are set to the on-site Coulomb repulsion $U = \SI{6}{\electronvolt}$ and the Hund's coupling $J_{\rm H} = \SI{0.7}{\electronvolt}$~\cite{Mizokawa1996}.
We have confirmed that the spin-orbital state obtained for MnV$_2$O$_4$ remains robust against variations in the Coulomb interaction parameters over the ranges considered (see Sec.~S5 of the SM).

The hopping matrix between neighboring V ions is constrained by the crystal symmetry of the spinel structure, in which the V ions form a pyrochlore lattice and are surrounded by edge-sharing octahedra, as shown in Fig.~\ref{fig:structure}a.
For the bond connecting sublattices 1 and 4, the hopping matrix is given by
\begin{equation}
T_{14} = \mqty( t_1 & t_2 & -t_4 \\ t_2 & t_1 & -t_4 \\ t_4 & t_4 & t_3 ), 
\end{equation}
expressed in the global $t_{2g}$ orbital basis $\{d_{yz}, d_{zx}, d_{xy}\}$. 
Here, $t_1$, $t_2$, $t_3$, and $t_4$ are hopping integrals.
The hopping matrices for the other bonds are obtained by symmetry operations.

\subsubsection*{Energy hierarchy and low-energy degrees of freedom}

We derive the one-body parameters from density functional theory (DFT) calculations for the high-temperature cubic phase of MnV$_2$O$_4$, including the hopping integrals $t_i$, the trigonal crystal field splitting $\Delta_{\rm tri}$, and the SOC constant $\zeta$ (see Methods for details).
The resulting parameters are summarized in Table~\ref{table:one-body_parameters}. 
We find that the hopping amplitude $t_3$ is dominant, leading to an exchange energy scale $J_0 = (t_3)^2 / U = \SI{9.2}{\milli\electronvolt}$. 
This scale is comparable to the effective SOC strength $\lvert \lambda \rvert = \zeta / (2S) = \SI{13.5}{\milli\electronvolt}$, indicating that SOC plays a non-negligible role. 
More notably, the trigonal crystal field splitting is significantly larger, $\Delta_{\rm tri} = \SI{42}{\milli\electronvolt}$, resulting in the hierarchy $\Delta_{\rm tri} > \lambda \sim J_0$. 
This hierarchy indicates that the trigonal crystal field plays the dominant role in determining the low-energy orbital manifold even in the cubic phase, providing the basis for constructing the effective spin-orbital model discussed below.

Under the trigonal crystal field, the threefold-degenerate $t_{2g}$ orbitals split into an $a_{1g}$ singlet and an $e_g^\pi$ doublet, as shown in Fig.~\ref{fig:structure}b. 
The $a_{1g}$ wave function is given by $\ket*{a_{1g}} = (\ket*{yz} +\ket*{zx} +\ket*{xy})/\sqrt{3}$ and its level lies lower in energy for $\Delta_{\rm tri} > 0$.
The $t_{2g}$ manifold carries an effective orbital angular momentum $l_{\rm eff}=1$, and the $a_{1g}$ and $e_g^\pi$ states correspond to the eigenstates of $l_{\rm eff}^Z=0$ and $l_{\rm eff}^Z=\pm1$, respectively, where the $Z$ axis is taken along the local trigonal axis, as shown in Fig.~\ref{fig:structure}a.
For the $d^2$ configuration of V$^{3+}$, one electron occupies the lower $a_{1g}$ orbital, while the other occupies one of the degenerate $e_g^\pi$ orbitals. 
Hund's coupling then favors the high-spin state with $S=1$. 
As a result, the low-energy manifold is described by the sixfold-degenerate states $\ket*{S=1,S^z,L_{\rm eff}^Z=\pm1}$, namely a spin triplet combined with a twofold orbital degeneracy. 
Here, $L_{\rm eff}^Z$ represents the total effective angular momentum of the two-electron state.

To describe the residual orbital degree of freedom within this low-energy manifold, we introduce an orbital pseudospin-$1/2$ operator $\tau^\mu$ ($\mu=X,Y,Z$) acting on the $\ket*{L_{\rm eff}^Z=\pm1}$ doublet within the two-electron $t_{2g}$ manifold. 
The $\tau^X$ and $\tau^Y$ eigenstates correspond to real orbitals and are even under time reversal, whereas the $\tau^Z$ eigenstate corresponds to a complex orbital and is odd under time reversal.
More explicitly, the remaining electron in the $e_g^\pi$ manifold can be expressed as linear combinations of the $t_{2g}$ orbitals. 
A convenient real-orbital basis is given by $\ket*{\psi_\alpha} = (\ket{yz}+\ket{zx}-2\ket{xy})/\sqrt{6}$ and $\ket*{\psi_\beta} = (\ket{yz}-\ket{zx})/\sqrt{2}$. 
The eigenstates of the effective orbital angular momentum are related to these real orbitals by $\ket*{l_{\rm eff}^Z=\pm1} = e^{\mp i\pi/6} (i\ket*{\psi_\alpha} \pm \ket*{\psi_\beta})/\sqrt{2}$, corresponding to the $\tau^Z=\pm1/2$ states.
Representative real and complex orbital wave functions are shown in Fig.~\ref{fig:structure}c.
Here, the above expressions are given for the sublattice whose trigonal axis is along $(1,1,1)/\sqrt{3}$; the corresponding expressions for the other sublattices are obtained by symmetry operations.

\begin{table}[t]
    \caption{
    \textbf{One-body parameters derived from DFT calculation.} 
		The one-body parameters include hopping integrals $t_i$, trigonal crystal-field splitting $\Delta_{\mathrm{tri}}$, and SOC $\zeta$. }
    \renewcommand{\arraystretch}{1.2}
    \label{table:one-body_parameters}
    \begin{ruledtabular}
        \begin{tabular}{ccccccc}
    		\quad $t_1$ & $t_2$ & $t_3$ & $t_4$ & $\Delta_{\mathrm{tri}}$ & $\zeta$ & (meV) \\[1mm] 
    		\quad $86$ & $-72$ & $-233$ & $-40$ & $42$ & $27$ &  \\ 
    	\end{tabular}
    \end{ruledtabular}
\end{table}

\begin{figure*}
\centering
\includegraphics[width=15cm]{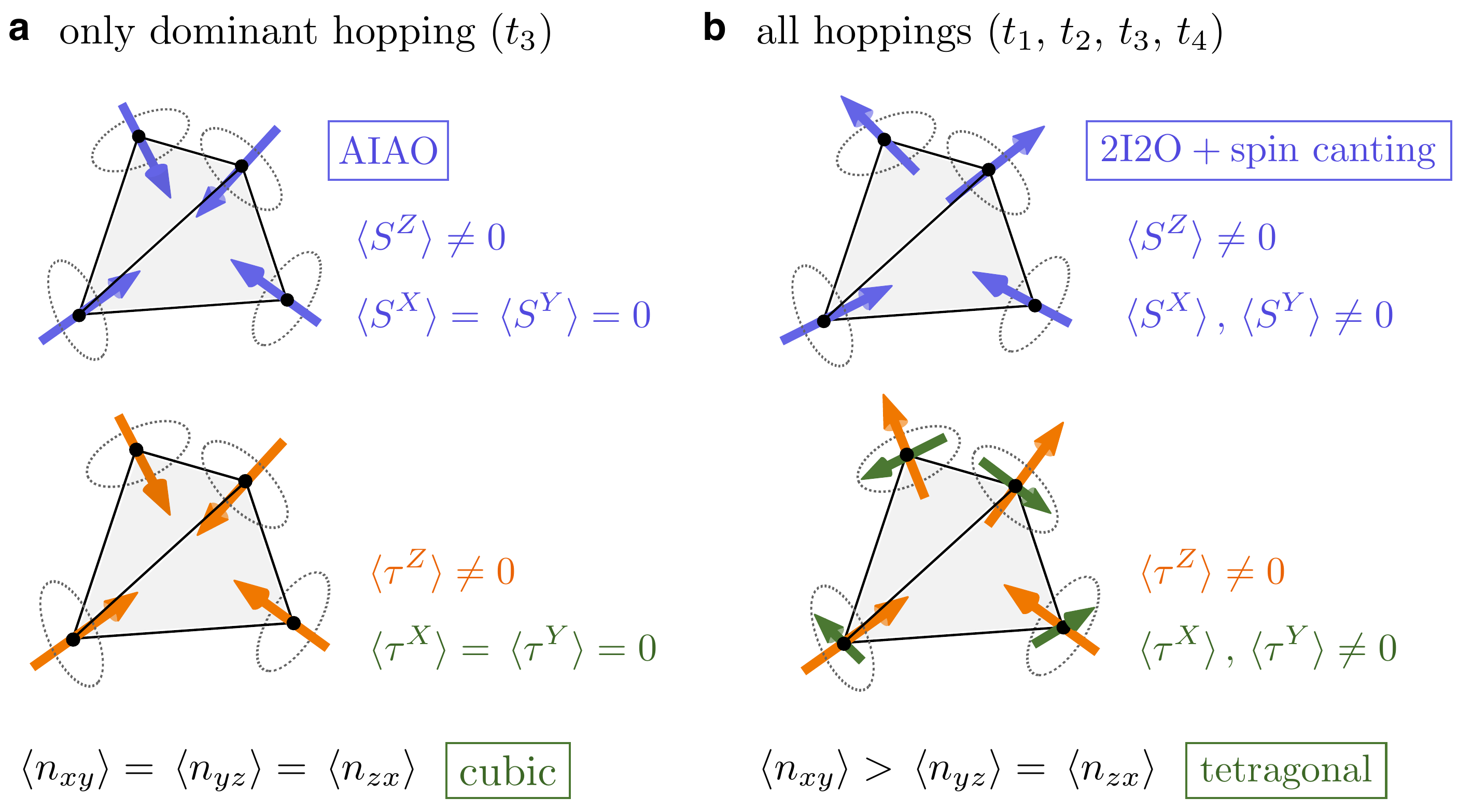}
\caption{
\textbf{Comparison between the model with only the dominant hopping and that with all hoppings.}
\textbf{a} Spin-orbital order obtained for only the dominant hopping ($t_3$). 
The spin sector exhibits an all-in/all-out (AIAO) configuration with $\ev*{S^Z} \neq 0$ and $\ev*{S^X} = \ev*{S^Y} = 0$. 
The orbital sector shows finite dipole order $\ev*{\tau^Z} \neq 0$ with vanishing quadrupole components $\ev*{\tau^X} = \ev*{\tau^Y} = 0$, resulting in uniform orbital occupations $\ev*{n_{xy}} = \ev*{n_{yz}} = \ev*{n_{zx}}$, compatible with cubic symmetry.
\textbf{b} Spin-orbital order obtained for all hoppings ($t_1$, $t_2$, $t_3$, $t_4$). 
The spin sector exhibits a two-in/two-out (2I2O) configuration with spin canting arising from finite transverse components $\ev*{S^X}, \ev*{S^Y} \neq 0$.
In the orbital sector, in addition to the dipole order, finite quadrupole components $\ev*{\tau^X}, \ev*{\tau^Y}$ emerge, leading to unequal orbital occupations $\ev*{n_{xy}} > \ev*{n_{yz}} = \ev*{n_{zx}}$, indicating an instability toward tetragonal symmetry.
}
\label{fig:ground_state}
\end{figure*}

\subsubsection*{Effective spin-orbital model}

Using degenerate perturbation theory, we derive an effective spin-orbital Hamiltonian within the low-energy manifold identified above. 
The unperturbed Hamiltonian is taken as $\mathcal{H}_0 = \mathcal{H}_{\rm C} + \mathcal{H}_{\rm CEF}^{\rm tri}$, and the perturbation as $\mathcal{V} = \mathcal{H}_{\rm SOC} + \mathcal{H}_{\rm kin}$. 
The perturbation is treated up to second order.
The detailed derivation is provided in Methods.
The validity of this low-energy description for the realistic parameters of MnV$_2$O$_4$ is examined in Sec.~S3 of the SM.

The effective Hamiltonian consists of two contributions, $\mathcal{H}_{\rm eff}=\mathcal{H}_{\rm ex}+\mathcal{H}_{\lambda}$, where $\mathcal{H}_{\rm ex}$ describes intersite spin-orbital exchange interactions generated by virtual hopping processes, and $\mathcal{H}_{\lambda}$ is an on-site term induced by SOC. 
The exchange term takes the form
\begin{equation}\label{eq:Ham_ex}
\begin{split}
\mathcal{H}_{\rm ex} &= \sum_{\langle i,j \rangle} \Big[ 
K_{zz} (1+\rho_{\parallel} \bar{\vb*{S}}_i\cdot\bar{\vb*{S}}_j) \tau_i^Z \tau_j^Z \\[-1mm]
&\quad\qquad - K_{\pm} (1+\rho_{\perp}\bar{\vb*{S}}_i\cdot\bar{\vb*{S}}_j) (\tau_i^+\tau_j^- +{\rm H.c.}) \\[1mm]
&\quad\qquad + K_{\pm\pm} (1+\rho_{\perp}\bar{\vb*{S}}_i\cdot\bar{\vb*{S}}_j) (\gamma^*_{ij}\tau_i^+\tau_j^+ +{\rm H.c.}) \\[1mm]
&\quad\qquad +J\bar{\vb*{S}}_i\cdot\bar{\vb*{S}}_j +\kappa \bar{\vb*{S}}_i\cdot\bar{\vb*{S}}_j (\gamma_{ij}[\tau_i^+ +\tau_j^+] +{\rm H.c.}) \Big], 
\end{split}
\end{equation}
and the on-site term is given by
\begin{equation}\label{eq:Ham_lam}
\begin{split}
\mathcal{H}_{\lambda}& = \sum_i \Big[ \lambda_\parallel \tau_i^Z S_i^Z +\lambda_\perp (\tau^X_i Q_i^{X^2-Y^2} +\tau_i^Y Q_i^{XY}) \\[-1mm]
&\quad\qquad +D_{z} Q_i^{3Z^2-R^2} \Big]. 
\end{split}
\end{equation}
Here, the orbital pseudospins are defined in the local coordinate frame associated with each trigonal axis. 
For the spin operators, both local and global frames are used. 
In the on-site term, the spin components $S_i^\mu$ ($\mu = X, Y, Z$) are defined in the local frame together with the spin quadrupole operators $Q_i^{X^2-Y^2}$, $Q_i^{XY}$, and $Q_i^{3Z^2-R^2}$ constructed as bilinears of $S=1$ operators. 
In contrast, $\bar{\vb*{S}}_i=(S_i^x,S_i^y,S_i^z)$ denotes the spin operator in the global frame used in the exchange term.
The bond-dependent phase factor $\gamma_{ij}$ takes the values $1$, $\omega$, and $\omega^2$ for the three inequivalent nearest-neighbor bonds labeled as $x$, $y$, and $z$, respectively, with $\omega = e^{2\pi i/3}$, following the convention in Ref.~\cite{Ross2011}. 
The $x$, $y$, and $z$ bonds are defined according to their orientations in the global coordinate frame, e.g., the $z$ bond is perpendicular to the global $z$ axis. 

Equation~(\ref{eq:Ham_ex}) describes bond-dependent spin-orbital exchange interactions. 
The first three terms represent bilinear orbital pseudospin interactions: the $K_{zz}$ term involves the orbital dipole component $\tau^Z$, whereas the $K_{\pm}$ and $K_{\pm\pm}$ terms involve the transverse components $(\tau^X,\tau^Y)$ corresponding to orbital quadrupoles. 
The fourth term is the Heisenberg exchange between the $S=1$ moments, and the last term proportional to $\kappa$ couples the spin correlation $\bar{\vb*{S}}_i \cdot \bar{\vb*{S}}_j$ to the orbital quadrupoles. 
In addition, $\rho_\parallel$ and $\rho_\perp$ are dimensionless parameters characterizing the ratios between the orbital exchange interactions and their spin-coupled counterparts, determined solely by the atomic parameters $U$, $J_{\rm H}$, and $\Delta_{\rm tri}$. 
In the present parameter set, they take the values $\rho_\parallel = 0.895$ and $\rho_\perp = 0.698$.
Within the second-order perturbation, $\mathcal{H}_{\rm ex}$ [Eq.~(\ref{eq:Ham_ex})] preserves SU(2) symmetry in spin space. Higher-order processes involving SOC may generate exchange interactions that break this symmetry.

Equation~(\ref{eq:Ham_lam}) indicates that SOC generates not only a coupling between the orbital dipole $\tau^Z$ and the spin dipole $S^Z$, but also couplings between the orbital quadrupoles and the spin quadrupole operators. 
The term proportional to $D_z$ describes a single-ion anisotropy of the $S=1$ moments. 

\subsection*{Magnetic and orbital structures and their microscopic mechanisms}

\subsubsection*{Intertwined dipole-quadrupole orbital order}

We investigate the ground-state properties of the effective Hamiltonian $\mathcal{H}_{\rm eff} = \mathcal{H}_{\rm ex} +\mathcal{H}_{\lambda}$ within a mean-field framework based on SU(N) coherent states~\cite{Nemoto2000}.
The detailed formulation, including the variational wave function and the definitions of the spin and orbital order parameters, is provided in Methods and Sec.~S4 of the SM. 

We first consider the limit where only the dominant hopping is retained, as it is expected to give the leading contribution to the exchange interactions, following previous theoretical studies~\cite{Tsunetsugu2003, Matteo2005, Chern2010}.
The corresponding model parameters, including exchange and on-site terms, are summarized in Table~\ref{table:model_parameters}. 
The obtained state exhibits an all-in/all-out (AIAO) magnetic structure together with orbital dipole order associated with $\tau^Z$, with no quadrupolar components ($\tau^X$ and $\tau^Y$). 
As a result, it yields uniform orbital occupations $\ev*{n_{xy}} = \ev*{n_{yz}} = \ev*{n_{zx}}$, as shown in Fig.~\ref{fig:ground_state}a, and therefore fails to account for the experimentally observed tetragonal distortion.

However, including subdominant hopping processes qualitatively changes the nature of the ground state. 
Instead of the AIAO state found in the dominant-hopping limit, the system stabilizes a two-in/two-out (2I2O) magnetic configuration with spin canting and intertwined dipole-quadrupole orbital order, as shown in Fig.~\ref{fig:ground_state}b. 
While the orbital dipole component $\ev*{\tau^Z}$ remains finite, additional quadrupolar components $\ev*{\tau^X}, \ev*{\tau^Y}\neq 0$ emerge, leading to unequal orbital occupations $\ev*{n_{xy}} > \ev*{n_{yz}} = \ev*{n_{zx}}$. 
This state is consistent with the experimentally observed tetragonal distortion, demonstrating that subdominant hopping processes play an essential role in stabilizing the spin-orbital order.

\subsubsection*{SOC-induced Ising anisotropy and noncollinear magnetic structure}

The magnetic structures obtained above are dominated by the $Z$ component: the AIAO state consists purely of $Z$ components, while the 2I2O state acquires additional transverse components but remains predominantly Ising-like. 
This Ising anisotropy originates from local effects arising from the interplay between the trigonal crystal field and the SOC~\cite{Yang2020}.
Although SOC is relatively small in $3d$ systems, it becomes important for understanding magnetism in $t_{2g}$ orbital systems with orbital degeneracy~\cite{Kanamori1957}.
The trigonal crystal field sets the local quantization axis on each sublattice, along which the dominant on-site SOC term $\lambda_\parallel$ favors alignment between the orbital dipole $\tau^Z$ and the spin component $S^Z$. 
Because the local $Z$ axes differ from site to site depending on the sublattice, this Ising anisotropy leads to a noncollinear magnetic structure in the global frame. 

The relation between the spin and orbital contributions to the magnetic moment can be understood from the expression $\vb*{\mu} = -\mu_{\rm B}\,(2\vb*{S} + \vb*{L}) = -\mu_{\rm B}\,(2\vb*{S} - \vb*{L}_{\rm eff})$, where $\vb*{L} = -\vb*{L}_{\rm eff}$ for the $t_{2g}$ manifold. 
The negative sign of $\lambda_\parallel$ favors parallel alignment between $\tau^Z$ and $S^Z$. 
Since $\tau^Z$ represents $L_{\rm eff}^Z$, the resulting orbital magnetic moment is antiparallel to the spin magnetic moment, leading to a reduction of the total magnetic moment. 

\begin{figure}
\centering
\includegraphics[width=7.5cm]{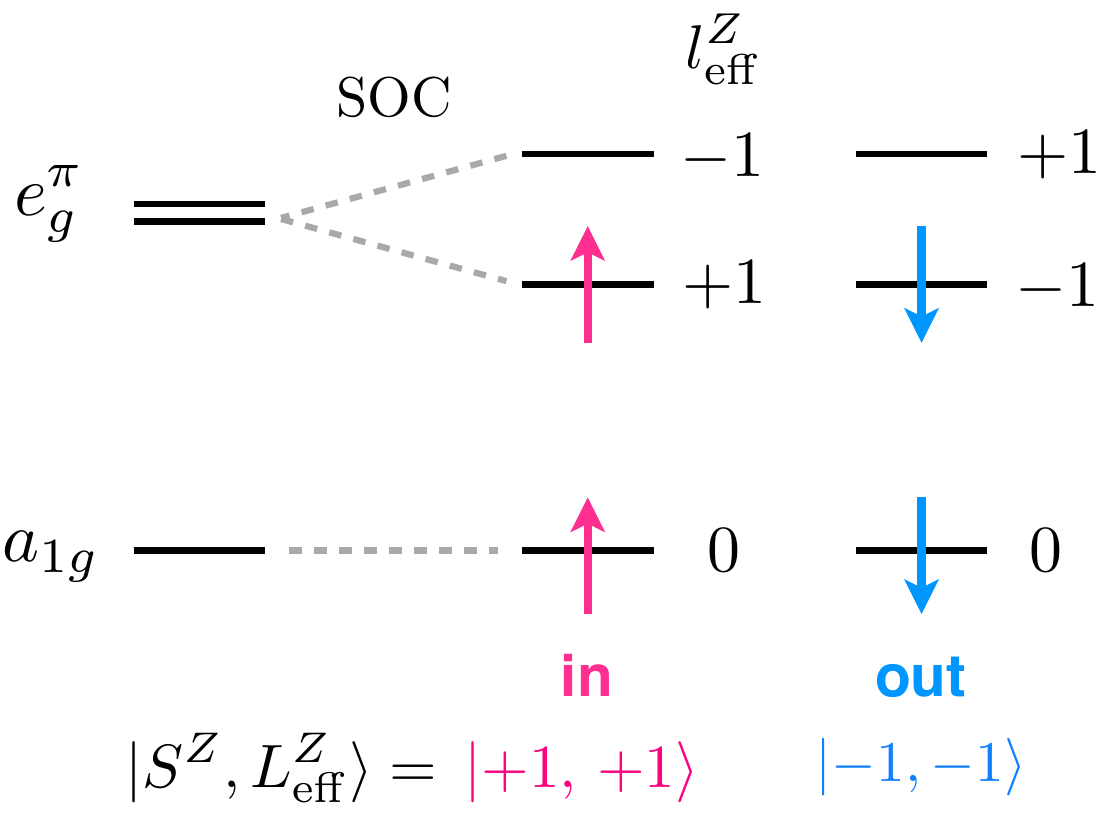}
\caption{
\textbf{Two-electron configurations under trigonal crystal field and SOC.}
Energy splitting of $e_g^\pi$ orbitals induced by SOC. 
In the resulting two-electron states, SOC selects $L_{\rm eff}^Z=\pm1$ states in the $e_g^\pi$ manifold, which are locked to $S^Z=\pm1$ through the $\lambda_{\parallel} \tau^Z S^Z$ coupling. 
These states define the local ``in'' and ``out'' configurations along the trigonal axes.
}
\label{fig:soc}
\end{figure}

In the strong SOC regime, the local states are described by $\ket*{S^Z=+1, L_{\rm eff}^Z=+1}$ and $\ket*{S^Z=-1, L_{\rm eff}^Z=-1}$, corresponding to the ``in'' and ``out'' states. 
These states can be expressed as two-electron configurations in the $a_{1g}$ and $e_g^\pi$ orbitals, where one electron occupies the $a_{1g}$ orbital and the other resides in the $e_g^\pi$ manifold. 
Within the $e_g^\pi$ manifold, SOC selects states with $l_{\rm eff}^Z = \pm 1$, such that the occupied orbital state is locked to the spin direction, as shown schematically in Fig.~\ref{fig:soc}. 
The exchange interactions then determine how these in/out moments are arranged across the lattice.

\begin{table}[t]
    \caption{
    \textbf{Model parameters of the effective Hamiltonian.}
	Exchange parameters [Eqs.~(\ref{eq:Ham_ex}) and (\ref{eq:Ham_lam})] obtained from the strong-coupling expansion are shown for two cases: 
    the model including only the dominant hopping $t_3$, and the model incorporating all hopping processes ($t_1$, $t_2$, $t_3$, and $t_4$) obtained from the DFT calculations. 
    The on-site parameters are evaluated for $\zeta=\SI{27}{\milli\electronvolt}$.}
    \renewcommand{\arraystretch}{1.2}
    \label{table:model_parameters}
    \begin{ruledtabular}
        \begin{tabular}{lccccc}
        Exchanges (meV) & $K_{zz}$ & $K_{\pm}$ & $K_{\pm\pm}$ & $J$ & $\kappa$ \\
        only dominant hopping ($t_3$) & $0$ & $-1.82$ & $1.82$ & $2.93$ & $-2.00$ \\
        all hoppings ($t_1$, $t_2$, $t_3$, $t_4$) & $18.2$ & $-10.4$ & $0.231$ & $2.64$ & $-1.42$ \\
        \hline
        On-site terms (meV) & $\lambda_\parallel$ & $\lambda_\perp$ & $D_z$ & & \\ 
        & $-22.7$ & $4.53$ & $1.20$ & & \\ 
        \end{tabular}
    \end{ruledtabular}
\end{table}

\begin{figure*}
\centering
\includegraphics[width=17cm]{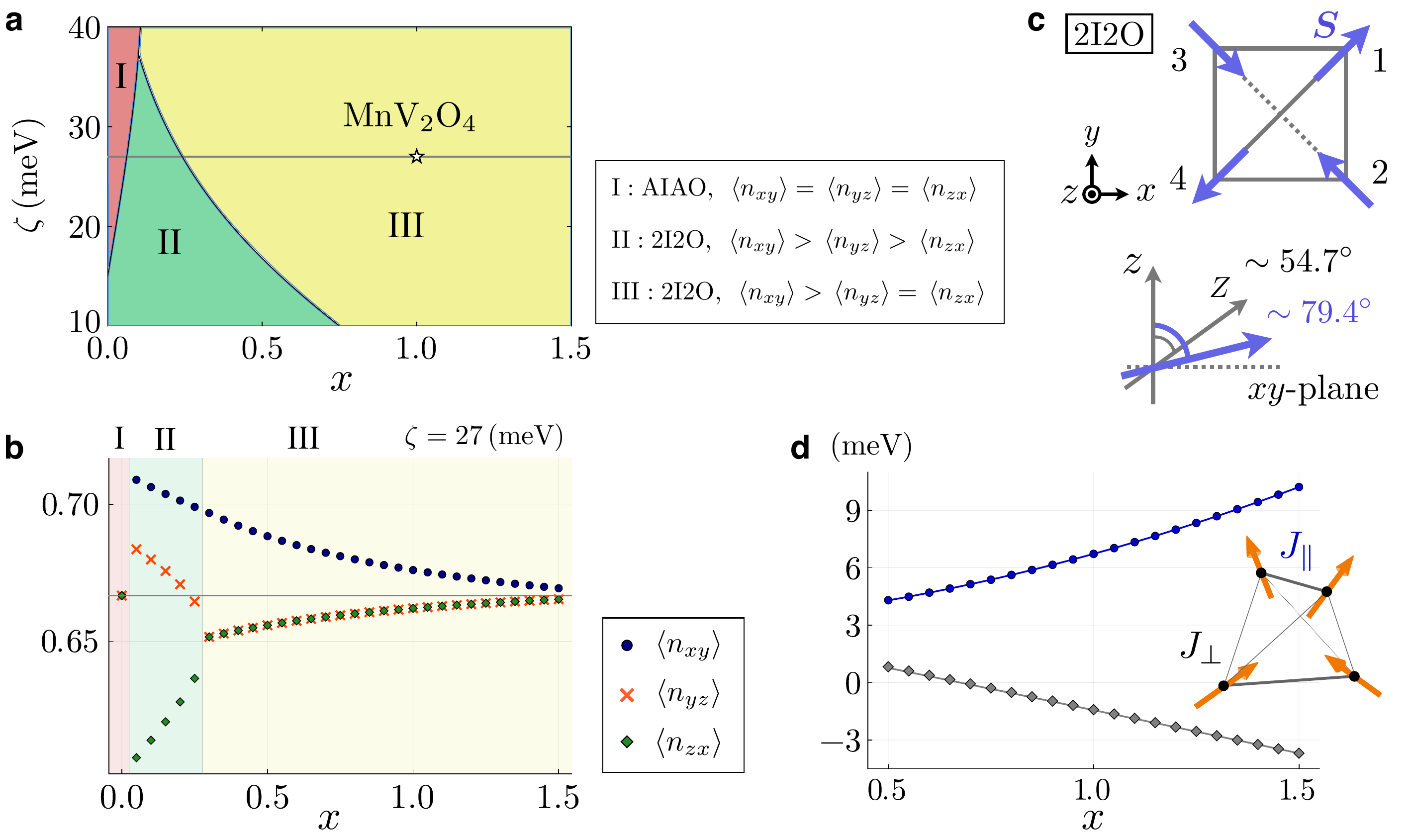}
\caption{
\textbf{Phase diagram, orbital occupations and magnetic structure.}
\textbf{a} Mean-field phase diagram in the $x$–$\zeta$ plane, where $x$ is the hopping tuning parameter and $\zeta$ is the SOC constant. 
\textbf{b} Orbital occupations $\ev*{n_m}$ ($m=yz,zx,xy$) in each phase as a function of $x$  at $\zeta = 27~\mathrm{meV}$.  
\textbf{c} Spin configuration in Phase~III, showing the noncoplanar structure and canting toward the global $xy$ plane.
The canting angle shown in the figure is the value obtained at the point corresponding to MnV$_2$O$_4$.
\textbf{d} Effective spin exchange constants $J_{\parallel}$ (blue) and $J_{\perp}$ (gray) as a function of $x$ in Phase~III under the 2I2O magnetic structure.
}
\label{fig:phase}
\end{figure*}

\subsubsection*{Mean-field phase diagram under subdominant hopping and SOC}

To elucidate how the inclusion of subdominant hopping processes modifies the ground state, we introduce a tuning parameter $x$ by decomposing the hopping matrix as $T(x)=T' + x T''$. 
Here, $T'$ contains only the dominant hopping $t_3$, and $T''$ includes the remaining symmetry-allowed hoppings $t_1$, $t_2$, and $t_4$. 
In this parametrization, $x=0$ corresponds to the dominant-hopping limit, while $x=1$ corresponds to the case where all hoppings are included.
In addition to $x$, the SOC constant $\zeta$ provides another key energy scale that competes with the exchange interactions. 
We therefore consider the $(x,\zeta)$ parameter space to clarify how the interplay between exchange interactions and SOC determines the ground state. 

Three distinct ordered phases are identified in this parameter space, as shown in Fig.~\ref{fig:phase}a. 
At $x=0$, the system is in the AIAO phase with uniform orbital occupations as we discussed above. 
Upon introducing a finite $x$, however, this phase is destabilized, and the system evolves through Phase~II into Phase~III as $x$ increases. 
MnV$_2$O$_4$ is located in Phase~III at $(x,\zeta)=(1,\SI{27}{\milli\electronvolt})$, well separated from the phase boundary, indicating that the realized order is robust against parameter variations. 

Both Phase~II and Phase~III exhibit the same longitudinal 2I2O configuration but differ in their transverse components, leading to distinct spin canting patterns and unequal orbital occupations.
In particular, the transverse components of the orbital pseudospin, $\tau^X$ and $\tau^Y$, corresponding to orbital quadrupole moments, give rise to unequal occupations of the $t_{2g}$ orbitals.
Figure~\ref{fig:phase}b shows the evolution of the orbital occupations across the three phases; Phase~I, with vanishing orbital quadrupole moments, exhibits uniform occupations, whereas finite quadrupole moments in Phases~II and III lead to unequal orbital occupations.

\subsubsection*{Origin of the two-in/two-out magnetic order and spin canting}
\label{sec:noncoplanar}

We now examine the microscopic mechanism underlying the formation of the magnetic structures.
We first consider the strong-SOC limit, in which the dominant on-site coupling $\lambda_{\parallel} \tau^Z S^Z$ constrains the low-energy degrees of freedom to Ising variables defined along the local trigonal axes, corresponding to the ``in'' and ``out'' states introduced above. 
Projecting the exchange Hamiltonian [Eq.~(\ref{eq:Ham_ex})] onto this manifold, the relevant interactions reduce to those involving the $Z$ components, i.e., the $K_{zz}$ and $J$ terms. 

Within this reduced description, the exchange energies per tetrahedron can be evaluated for the AIAO and 2I2O configurations:
\begin{equation}\label{eq:Ising}
\begin{split}
& E_{\rm AIAO} = \frac{3-\rho_\parallel}{4} K_{zz} -J, \\[2mm]
& E_{\rm 2I2O} = -\frac{1+\rho_\parallel}{4} K_{zz} +\frac{1}{3}J, 
\end{split}
\end{equation}
where $\rho_\parallel = 0.895$ as shown above. 
These expressions show that the $K_{zz}$ term favors the 2I2O configuration, whereas the $J$ term stabilizes the AIAO state. 
When only the dominant hopping is retained, $K_{zz}=0$ (see Table~\ref{table:model_parameters}), and the AIAO state is realized.  
In contrast, when all hoppings are included, $K_{zz}$ becomes the largest exchange parameter, with $K_{zz}/J \sim 7$, strongly favoring the 2I2O state relevant to MnV$_2$O$_4$ in the present parameter regime.

Having established the longitudinal 2I2O configuration, we next consider the transverse spin components and the resulting canting of spins. 
The spins are tilted within the plane defined by the global $z$ axis and each local $Z$ axis, toward the global $xy$ plane, as shown in Fig.~\ref{fig:phase}c. 
The canting angle between the spins and the global $z$ axis is estimated to be approximately $79^\circ$ for the parameter set corresponding to MnV$_2$O$_4$. 
This value is likely overestimated within the present framework, as will be discussed in the Discussion section.

To elucidate the origin of this canting, we examine the effect of orbital order on the spin interactions. 
By replacing the orbital pseudospin operators with their expectation values, the exchange becomes bond dependent and can be described by an effective spin Hamiltonian,
\begin{equation}
\mathcal{H}_{\rm spin} = \sum_{\langle i,j \rangle} J_{ij} \, \bar{\vb*{S}}_i \cdot \bar{\vb*{S}}_j,
\end{equation}
where the exchange constants depend on the underlying orbital configuration. 
In the 2I2O state, the interactions split into two types,
\begin{equation}
J_{\parallel} = J + \frac{\rho_\parallel}{4} K_{zz}, \quad
J_{\perp} = J - \frac{\rho_\parallel}{4} K_{zz},
\end{equation}
corresponding to bonds connecting sites with the same or opposite orbital orientations, respectively. 
As shown in Fig.~\ref{fig:phase}d, $J_{\parallel}$ is enhanced by the $K_{zz}$ term and becomes strongly antiferromagnetic on bonds of the former type, such as those lying in the $xy$ plane.
This strong antiferromagnetic $J_{\parallel}$ tends to align spins antiparallel on the corresponding bonds.
As a result, the spins tilt toward the global $xy$ plane, giving rise to the noncoplanar magnetic structure observed in Phase~III.
We briefly note that in Phase~II the spins exhibit an additional in-plane rotation, along with the canting toward the $xy$ plane, which further lowers the symmetry of the ordered state.
The explicit spin configurations in Phases~II and III are presented in Sec.~S4 of the SM.

\subsubsection*{Instability toward tetragonal compression}

The preceding analysis shows that the spin canting originates from spin exchange interactions modulated by the orbital dipole order. 
At the same time, the orbital quadrupole moments become finite in the 2I2O state, which modifies the occupation of the $t_{2g}$ orbitals. 
These quadrupole moments couple to the lattice degrees of freedom and can drive a structural distortion. 
In the following, we elucidate the microscopic origin of these quadrupole moments.

\begin{figure}
\centering
\includegraphics[width=\linewidth]{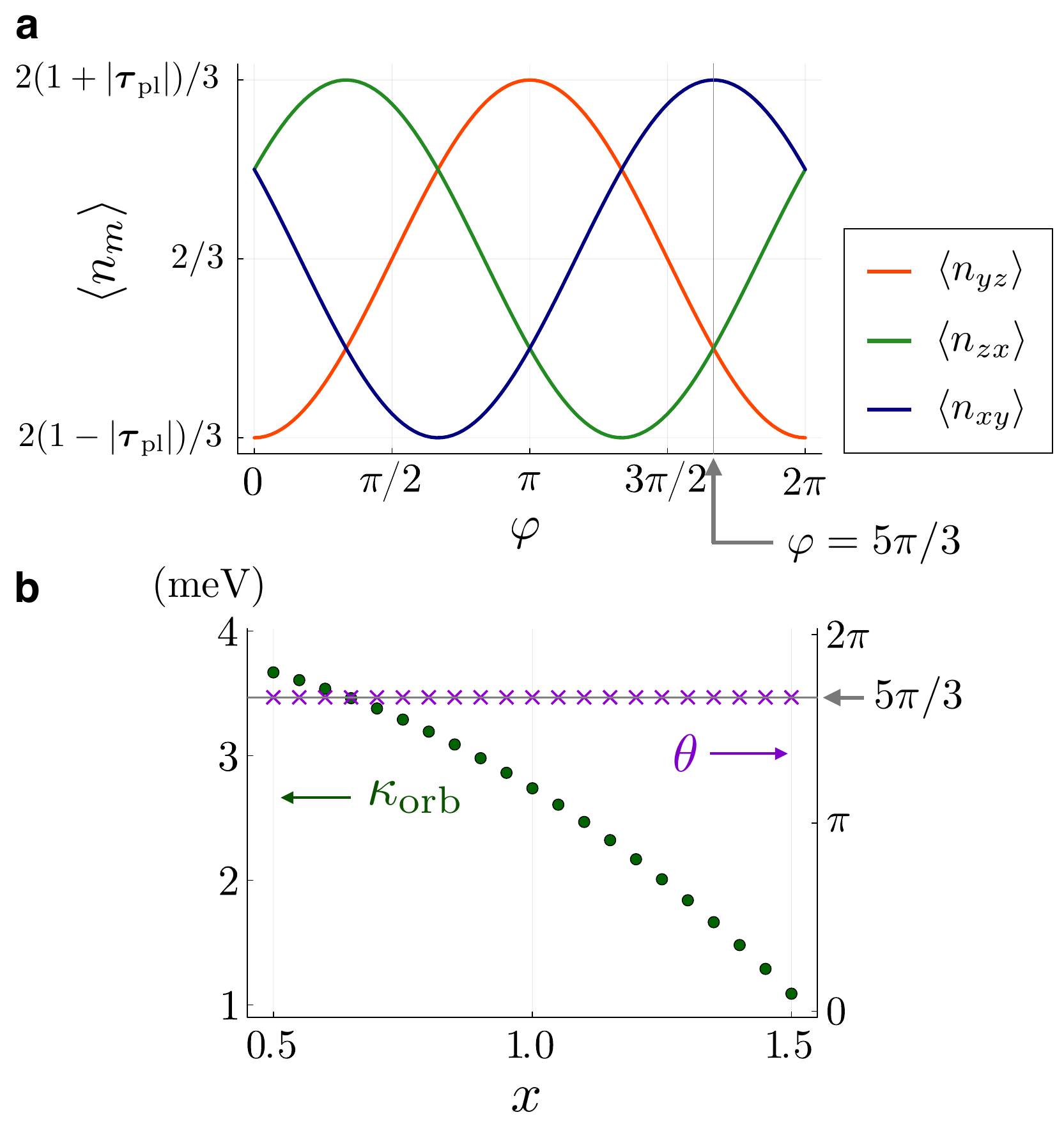}
\caption{
\textbf{Orbital occupations and effective quadrupolar field.}
\textbf{a} Orbital occupations $\ev*{n_m}$ ($m=yz,zx,xy$) as a function of the angle $\varphi$ that parametrizes the transverse orbital pseudospin $\vb*{\tau}_{\rm pl}$, defined by $(\ev*{\tau^X}, \ev*{\tau^Y}) = |\vb*{\tau}_{\rm pl}|(\cos\varphi,\sin\varphi)$.
\textbf{b} Effective orbital field $\kappa_{\rm orb}$ and its direction $\theta$ in Phase~III as a function of the hopping tuning parameter $x$. 
}
\label{fig:orbital_occupation}
\end{figure}

The orbital occupations can be expressed using the orbital quadrupoles as
\begin{equation}\label{eq:orbital_occupations}
n_m=\frac{2}{3}\,\Big[1 - (\cos\theta_m \tau^X + \sin\theta_m \tau^Y)\Big],
\end{equation}
where $m=yz,zx,xy$ and $\theta_m = 0, -2\pi/3, +2\pi/3$.
This expression shows that the direction of the orbital pseudospin in the $\tau^X$-$\tau^Y$ plane determines the occupations of the three $t_{2g}$ orbitals. 
To visualize this relation, we parametrize the transverse orbital pseudospin $\vb*{\tau}_{\rm pl} = (\tau^X,\tau^Y)$ as $\ev*{\tau^X}=|\vb*{\tau}_{\rm pl}|\cos\varphi$ and $\ev*{\tau^Y}=|\vb*{\tau}_{\rm pl}|\sin\varphi$, and show the resulting $\varphi$ dependence of $\ev*{n_m}$ in Fig.~\ref{fig:orbital_occupation}a.

\begin{figure*}[t]
\centering
\includegraphics[width=16cm]{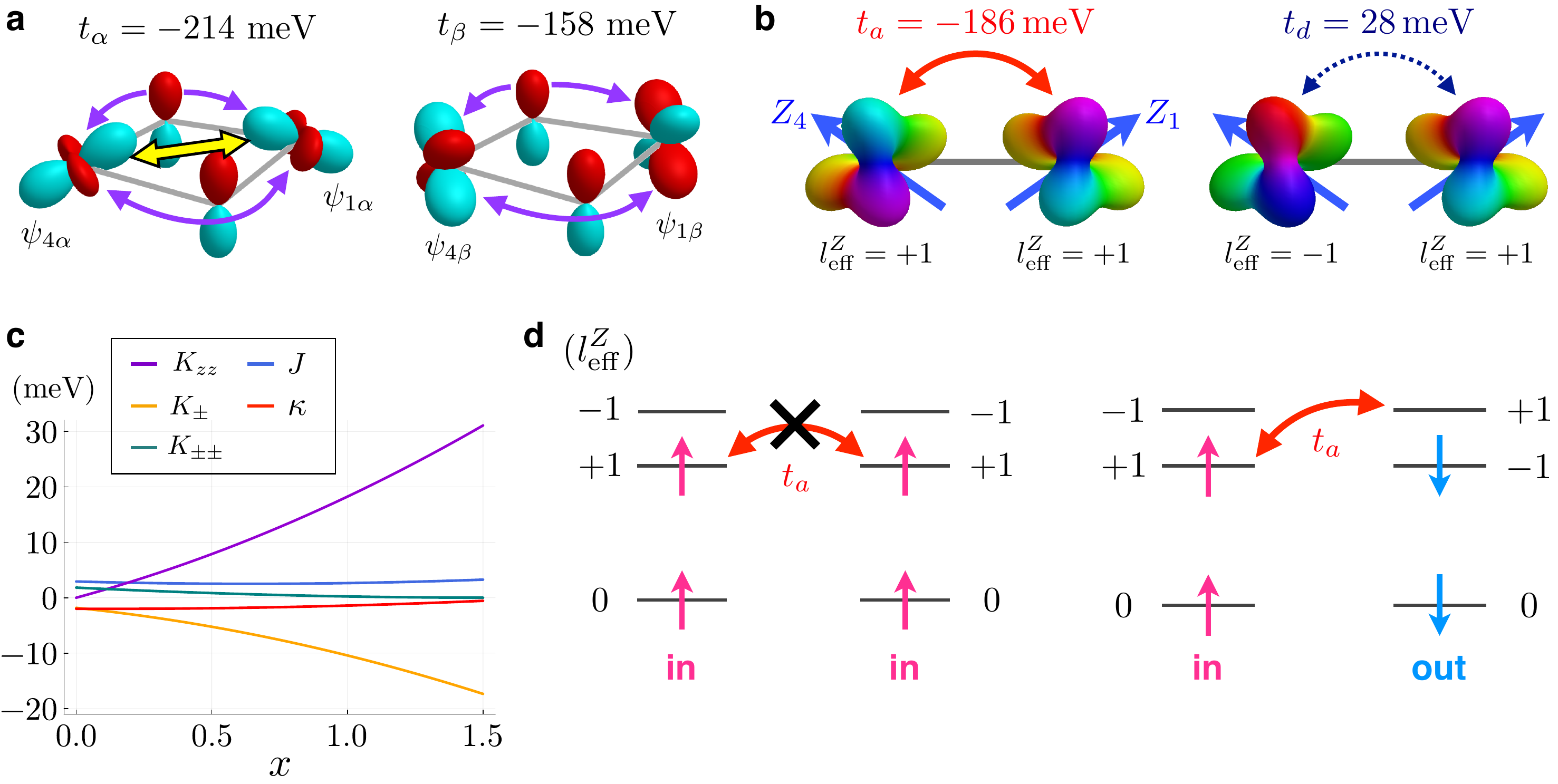}
\caption{
\textbf{Hopping processes and exchange interactions.}
\textbf{a,b} Schematic illustrations of hopping processes within the $e_g^\pi$ manifold in the real-orbital basis $\{\psi_\alpha,\psi_\beta\}$ (\textbf{a}) and the $l_{\rm eff}^Z=\pm1$ basis (\textbf{b}).
The hopping integrals $t_a$ and $t_d$ in the latter basis correspond to diagonal and off-diagonal processes, connecting states with the same and opposite $l_{\rm eff}^Z$, respectively.
The values shown correspond to the case with all hoppings included ($x=1$).
\textbf{b} Exchange parameters as a function of the tuning parameter $x$ in the hopping matrix. 
\textbf{c} Schematic illustration of the mechanism by which dominant diagonal hopping $t_a$ favors the 2I2O configuration. 
}
\label{fig:exchange}
\end{figure*}

We now examine how such quadrupole moments are induced by the spin configuration. 
By replacing the spin operators with their expectation values, the $\kappa$ term in $\mathcal{H}_{\rm ex}$, which represents the coupling term between spin correlations and orbital quadrupoles, becomes
\begin{equation}
\begin{split}
\mathcal{H}_{\kappa} &= \kappa \sum_{\langle i,j \rangle} \ev*{\bar{\vb*{S}}_i\cdot\bar{\vb*{S}}_j} \left( \gamma_{ij}[\tau_i^+ + \tau_j^+] + \mathrm{H.c.} \right) \\
&= -\sum_i \vb*{\kappa}_{i, \rm orb} \cdot \vb*{\tau}_i, 
\end{split}
\end{equation}
where $\vb*{\kappa}_{i,\rm orb}$ acts as an effective field on the orbital quadrupoles. 
We parametrize it as $\vb*{\kappa}_{i,\rm orb}=\kappa_{i,\rm orb}(\cos\theta_i,\sin\theta_i)$ with $\kappa_{i,\rm orb}\geq0$. 
Its magnitude and direction are determined by the underlying spin configuration.

This effective field vanishes in the AIAO configuration, as the contributions from neighboring sites exactly cancel by symmetry, leading to uniform orbital occupations.
In contrast, the 2I2O configuration generates a finite $\vb*{\kappa}_{i,\rm orb}$. 
For a 2I2O configuration with identical Ising pairs on a given $m$-plane, the direction of the effective field is given by $\theta_i=\theta_m$ or $\theta_m+\pi$, where $\theta_m$ is defined in Eq.~(\ref{eq:orbital_occupations}). 
Moreover, the effective field is uniform across all sublattices, leading to a ferroquadrupolar orbital order. 

The behavior of $\kappa_{\rm orb}$ and $\theta$ in Phase~III is shown in Fig.~\ref{fig:orbital_occupation}b. 
In Phase~III, including $x=1$ corresponding to the parameter set of MnV$_2$O$_4$, the angle $\theta$ is pinned to $\theta_{xy}+\pi$, leading to the orbital occupation pattern $\ev*{n_{xy}} > \ev*{n_{yz}}=\ev*{n_{zx}}$. 
This occupation pattern is consistent with the experimentally observed tetragonal compression in MnV$_2$O$_4$, and corresponds to the space group $I4_1/amd$. 
In contrast, in Phase~II, the additional in-plane rotation discussed above shifts $\theta_i$ away from $\theta_{xy}+\pi$, resulting in $\ev*{n_{xy}} > \ev*{n_{yz}} > \ev*{n_{zx}}$. 
Within the present framework, no orbital occupation pattern corresponding to the alternative space group $I4_1/a$ is obtained.


\subsection*{Impact of subdominant hoppings on spin-orbital exchange interactions}

As shown above, including or neglecting subdominant hopping processes leads to qualitatively different solutions, namely 2I2O and AIAO configurations, respectively. In particular, the emergence of a dominant $K_{zz}$ term upon incorporating these subdominant hopping contributions plays a crucial role in stabilizing the 2I2O configuration over the AIAO state (see Table~\ref{table:model_parameters} and Fig.~\ref{fig:ground_state}).
Under this 2I2O magnetic structure, additional features such as spin canting and ferroquadrupolar orbital order emerge, as discussed in the preceding analysis. 
In the following, we elucidate how the underlying hopping processes lead to the enhancement of the key interaction $K_{zz}$.

To clarify how the subdominant hopping processes modify the exchange interactions, we first consider the hopping processes in terms of the real orbitals $\psi_\alpha$ and $\psi_\beta$.
The hopping matrix within the $e_g^\pi$ manifold is diagonal in this basis, with hopping amplitudes $t_\alpha$ and $t_\beta$ for the $\psi_\alpha$ and $\psi_\beta$ orbitals, respectively, as shown in Fig.~\ref{fig:exchange}a.
Schematically, $t_\alpha$ involves both direct $d$-$d$ and oxygen-mediated hopping processes, whereas $t_\beta$ is predominantly associated with the oxygen-mediated process (see Sec.~S2 of the SM for details).
In the dominant-hopping limit, $t_\beta$ is neglected, whereas it becomes sizable when the subdominant hoppings are taken into account.
For the DFT-derived hopping parameters, we obtain $t_\alpha=-214$~meV and $t_\beta=-158$~meV.

While the real-orbital representation provides a transparent picture of the underlying hopping processes, the complex $l_{\rm eff}^Z=\pm1$ basis is more natural for describing the low-energy states in the presence of SOC.
The hopping amplitudes in the $l_{\rm eff}^Z$ basis are given by $t_a=(t_\alpha+t_\beta)/2$ and $t_d=(t_\beta-t_\alpha)/2$, where $t_a$ and $t_d$ connect states with the same and opposite $l_{\rm eff}^Z$, respectively, as shown in Fig.~\ref{fig:exchange}b.
Since $t_\alpha$ and $t_\beta$ have comparable magnitudes and the same sign, their contributions add in $t_a$ but largely cancel in $t_d$.
For MnV$_2$O$_4$, this gives $t_a=-186$~meV and $t_d=28$~meV, resulting in a substantial difference between $|t_a|$ and $|t_d|$ when the subdominant hoppings are included, in contrast to the dominant-hopping limit where $|t_a|=|t_d|$ (see Eqs.~(S20) and (S24)).


This contrast between diagonal and off-diagonal hopping processes is manifested in the exchange parameters. 
From the strong-coupling expansion, we obtain
\begin{equation}
\begin{split}
& K_{zz} = \frac{4[(t_a)^2-(t_d)^2]}{(1+\rho_\parallel)(U-3J_{\rm H})}, \\[2mm]
& K_{\pm} = -\frac{2(t_a)^2}{(1+\rho_\perp)(U-3J_{\rm H})}, \\[2mm]
& K_{\pm\pm} = \frac{2(t_d)^2}{(1+\rho_\perp)(U-3J_{\rm H})}. 
\end{split}
\end{equation}
In the dominant-hopping limit where $|t_a|=|t_d|$, one finds $K_{zz}=0$ and $K_{\pm\pm}=-K_{\pm}$, reducing the bilinear orbital interaction to a bond-dependent Ising form known as the compass model~\cite{Chern2010}. 
In contrast, the difference between $|t_a|$ and $|t_d|$ gives rise to a finite $K_{zz}$, driving the system away from the compass-model limit.
The evolution of the exchange parameters as a function of the hopping tuning parameter $x$ is shown in Fig.~\ref{fig:exchange}b. 
Upon including subdominant hoppings, the exchange interactions are qualitatively modified: $K_{zz}$ is rapidly enhanced even for small values of $x$, while $K_{\pm\pm}$, $J$, and $\kappa$ depend only weakly on $x$.


As an alternative to the discussion based on the $K_{zz}$ term [see Eq.~(\ref{eq:Ising})], one can also understand why the 2I2O configuration is favored in the presence of subdominant hopping by considering the substantial difference between $|t_a|$ and $|t_d|$ in the $l_{\rm eff}^Z$ basis (Fig.~\ref{fig:exchange}d).
The dominant diagonal hopping $t_a$
connects orbitals with the same $l_{\rm eff}^Z$, as illustrated schematically in Fig.~\ref{fig:exchange}a.
For bonds connecting two identical configurations (both in or both out), the relevant hopping processes are prohibited by the Pauli exclusion principle, since the corresponding spin-orbital states are already occupied. 
In contrast, for bonds connecting opposite configurations (in and out), the hopping process remains allowed, leading to a finite energy gain. 
As a result, configurations that maximize the number of in-out bonds are energetically favored, which stabilizes the 2I2O state.


\section*{Discussion}

We first discuss the relation between the present results and the previous study by Chern \textit{et al.}~\cite{Chern2010}, which considers the same low-energy manifold formed under the strong trigonal crystal field. 
In that study, only the dominant hopping $t_3$ is taken into account, leading to the compass model, which stabilizes an antiferroquadrupolar orbital order with the space group $I4_1/a$. 
The magnetic structure is proposed to be the 2I2O configuration, driven by the antiferromagnetic interaction between Mn and V ions. 
Furthermore, it is proposed that the transverse spin-orbital coupling term, corresponding to the $\lambda_{\perp}$ term in our notation [Eq.~(\ref{eq:Ham_lam})], reproduces the magnetic structure determined under the $I4_1/a$ symmetry~\cite{Garlea2008_MnV2O4}.

In contrast to this picture, our results demonstrate that the inclusion of subdominant hoppings qualitatively modifies the exchange interactions, resulting in a different orbital order and space-group symmetry. 
In particular, a finite $K_{zz}$ interaction, which couples orbital dipoles through a spin-dependent prefactor, is generated, which is absent in the compass limit and becomes the dominant exchange term, stabilizing the 2I2O configuration. 
Within this 2I2O state, the $\kappa$ term, which induces orbital quadrupole moments via spin correlations, further induces the ferroquadrupolar orbital order, leading to the $I4_1/amd$ symmetry instead of $I4_1/a$. 
Notably, this term remains finite even in the dominant-hopping limit and was effectively neglected in the previous study.
These results demonstrate that going beyond the compass-model limit by incorporating subdominant hoppings and all symmetry-allowed exchange interactions is essential for capturing the competing interactions that stabilize complex spin-orbital orders.

We next discuss the robustness and limitations of the present description.
To examine the stability of the spin-orbital state against quantum fluctuations, we perform linear flavor-wave calculations around the variational ground state. 
The flavor-wave excitation spectrum is gapped, and the quantum corrections to the spin, orbital-dipole, and orbital-quadrupole moments are moderate, on a scale of approximately 10\%. 
These moderate quantum corrections are consistent with the pronounced Ising-like character of the local spin-orbital states, which suppresses quantum fluctuations.
Moreover, the orientations of the spin and orbital moments remain unchanged. 
These results indicate that the 2I2O spin-orbital state is robust against quantum fluctuations. 
Details of the linear flavor-wave calculations are provided in Methods and Sec.~S6 of the SM.


We turn our attention to the spin canting angle.
The orbital magnetic moment is likely overestimated within the mean-field approximation, which affects the quantitative value of the canting angle.
A reduction of the orbital moment would weaken the anisotropy of the Heisenberg interactions between V ions, suppressing the tendency toward in-plane canting.
The Mn-V exchange interaction, which is not explicitly included in the present model, also affects the canting angle.
Our DFT calculations estimate this interaction to be antiferromagnetic and of the order of a few meV.
As shown by the symmetry-based analysis in Sec.~S4 of the SM, the collinear ferrimagnetic state favored in the strong Mn-V coupling limit and the noncollinear state obtained in the present model are described by the same two magnetic order parameters and differ only in their relative coefficients. 
The Mn-V interaction is therefore expected to change their relative weights and hence the spin canting angle.
Taken together, these effects are expected to reduce the canting angle and bring the spins closer to the global $z$ direction.

Finally, the Jahn-Teller coupling is not explicitly included in the present model. 
At an effective level, integrating out lattice degrees of freedom generates orbital-quadrupole interactions~\cite{Chern2010} of the same form as those already present in Eq.~(\ref{eq:Ham_ex}), providing corrections to the corresponding exchange parameters. 
Such corrections are not expected to qualitatively alter our conclusions unless the electron-lattice coupling is considerably strong, in which case the lattice degrees of freedom should be retained explicitly.

In summary, we have developed a microscopic understanding of the orbital and magnetic structures in MnV$_2$O$_4$ by constructing and analyzing an effective model based on DFT calculations for the high-temperature cubic phase. 
Our analysis demonstrates that the trigonal crystal field is an essential ingredient in determining the low-energy degrees of freedom, and that subdominant hopping processes strongly modify the spin-orbital exchange interactions beyond the conventional dominant-hopping picture. 
A key aspect of our approach is the quantitative evaluation of the hierarchy of interaction energy scales based on parameters extracted from DFT calculations, which is crucial for understanding spin-orbital physics in transition-metal compounds with multiple competing energy scales. 
Within this framework, the effective model stabilizes the 2I2O configuration of spin and orbital dipolar moments accompanied by spin canting and ferroquadrupolar orbital order. 
These results clarify how competing and cooperating exchange interactions govern the intertwined spin-orbital state and the associated instability toward tetragonal distortion, while highlighting how different microscopic hopping processes can interfere after projection onto low-energy orbital states selected by crystal fields and SOC, reshaping the exchange interactions in multiorbital systems. 

\section*{Methods}

\subsection*{DFT calculation}

To derive the one-body parameters, we perform DFT calculations for MnV$_2$O$_4$ using Quantum~ESPRESSO~\cite{Giannozzi2017} with inclusion of SOC. 
The calculations are performed for the high-temperature cubic structure provided by the experimental group of Ref.~\cite{Koyama2026}. 
Relativistic norm-conserving pseudopotentials with the Perdew-Burke-Ernzerhof (PBE) exchange-correlation functional~\cite{Perdew1996} are adopted from the PseudoDojo library~\cite{VanSetten2018}. 
Based on the DFT electronic structure, we construct a tight-binding model using RESPACK~\cite{Nakamura2021, Charlebois2021} and Wannier90~\cite{Pizzi2020}, from which the hopping parameters, trigonal crystal-field splitting, and SOC constant are extracted. 
The plane-wave energy cutoffs are set to 100~Ry for the wavefunctions and 400~Ry for the charge density, and a $6 \times 6 \times 6$ ${\boldsymbol k}$-mesh is employed. 
The resulting parameters are listed in Table~\ref{table:one-body_parameters}.

\subsection*{Strong-coupling expansion}

To obtain the effective spin-orbital Hamiltonian $\mathcal{H}_{\rm eff}$ [Eqs.~(\ref{eq:Ham_ex}) and (\ref{eq:Ham_lam})], we perform a strong-coupling expansion by treating $\mathcal{V} = \mathcal{H}_{\rm kin} + \mathcal{H}_{\rm SOC}$ as a perturbation to $\mathcal{H}_0 = \mathcal{H}_{\rm C} + \mathcal{H}_{\rm CEF}^{\rm tri}$.
Applying degenerate perturbation theory~\cite{Lindgren1974} up to second order in $\mathcal{V}$, we obtain
\begin{equation}
\begin{split}
\mathcal{H}_{\rm eff} &= \mathcal{P} \mathcal{V} \mathcal{P} + \mathcal{P} \mathcal{V} \frac{1}{E-\mathcal{H}_0} (\hat{1}-\mathcal{P})\mathcal{V}\mathcal{P} \\[1mm]
&= \mathcal{P} \mathcal{H}_{\rm SOC} \mathcal{P}
+ \mathcal{P} \mathcal{H}_{\rm SOC} \frac{1}{E-\mathcal{H}_0} (\hat{1}-\mathcal{P}) \mathcal{H}_{\rm SOC} \mathcal{P} \\
&\quad +\mathcal{P} \mathcal{H}_{\rm kin} \frac{1}{E-\mathcal{H}_0} \mathcal{H}_{\rm kin} \mathcal{P}. 
\end{split}
\end{equation}
Here, $\mathcal{P}$ denotes the projection operator onto the ground multiplet manifold spanned by $\ket*{S = 1, S^z, L_{\rm eff}^Z = \pm 1}$, and $E$ represents the corresponding multiplet energy.
The first two terms arise from the SOC and correspond to the on-site interaction $\mathcal{H}_{\lambda}$ [Eq.~(\ref{eq:Ham_lam})]; the first-order contribution generates $\lambda_\parallel$, while the second-order contribution involving excited multiplets provides a correction to $\lambda_\parallel$ and generates the $\lambda_\perp$ and $D_z$ terms.
The third term originates from virtual hopping processes and gives rise to the spin-orbital exchange interaction $\mathcal{H}_{\rm ex}$ [Eq.~(\ref{eq:Ham_ex})].

\subsection*{Mean-field analysis}

The effective Hamiltonian is analyzed within a mean-field framework based on a site-factorized variational wave function. 
The local quantum state is expressed as a direct product of spin and orbital coherent states~\cite{Nemoto2000},
\begin{equation}
\ket{\Psi} = \prod_i \ket*{\Omega_i^{\rm spin}} \otimes \ket*{\Omega_i^{\rm orb}},
\end{equation}
where $\ket*{\Omega_i^{\rm spin}}$ represents an SU(3) coherent state for the spin-$S=1$ degree of freedom, and $\ket*{\Omega_i^{\rm orb}}$ represents an SU(2) coherent state for the orbital pseudospin-$1/2$.
This variational ansatz captures local multipolar degrees of freedom while neglecting intersite quantum fluctuations. 

For the nearest-neighbor Hamiltonian on the pyrochlore lattice, the ground state can be obtained by minimizing the energy within a single tetrahedron~\cite{HanYan2017}.
The ordered phases are characterized by spin and orbital order parameters defined as expectation values of the corresponding operators. 
The explicit definitions and their classification based on the irreducible representations of the $T_d$ point group are given in Sec.~S4 of the SM.

The ground state is obtained by minimizing the variational energy with respect to the coherent-state parameters. 
To construct the phase diagram, we vary the hopping tuning parameter $x$ in the hopping matrix and the SOC constant $\zeta$, in addition to evaluating the model at the DFT-derived parameter set.

Within the present variational ansatz, the spin and orbital degrees of freedom are treated as separable, neglecting possible spin-orbital entanglement at the single-site level. 
We have verified that including spin-orbital entanglement in the local variational space does not qualitatively change the resulting phase diagram or the nature of the ordered phases. 

\subsection*{Linear flavor-wave theory}

To estimate the quantum correction to the ordered moments, we perform a linear flavor-wave analysis~\cite{Joshi1999, Lauchli2006, GangChen2010}.
For each sublattice, the local Hilbert space is transformed to a basis in which the mean-field state is chosen as the reference state, and fluctuations orthogonal to it are represented by bosons.
Expanding the Hamiltonian in these bosonic operators and retaining terms up to quadratic order yields the linear flavor-wave Hamiltonian, which is diagonalized by a bosonic Bogoliubov transformation.
The resulting boson occupations are used to evaluate the quantum corrections to the spin and orbital moments.

We perform the analysis using two choices of the local state.
First, consistently with the mean-field analysis above, we use a separable $\mathrm{SU}(3)\otimes\mathrm{SU}(2)$ state for the spin and orbital degrees of freedom and retain the corresponding spin and orbital fluctuation modes.
As a complementary calculation, we also determine the mean-field state allowing a general SU(6) local state, including on-site spin-orbital entanglement, and perform the flavor-wave expansion.
In this case, all local fluctuations orthogonal to the reference state are retained.
The two calculations provide complementary estimates of the quantum corrections to the ordered moments, with the detailed numerical results presented in Sec.~S6 of the SM.

\section*{Data availability}
The data that support the findings of this study are available from the corresponding authors upon reasonable request.

\section*{Code availability}
The code that supports the findings of this study is available from the corresponding authors upon reasonable request.

\bibliography{refs}

@article{Kugel1973,
	title = {{Crystal structure and magnetic properties of substances with orbital degeneracy}},
	author = {Kugel, K.~I. and Khomskii, D.~I.},
	journal = {Sov. Phys. JETP},
	year = {1973},
	month = {Oct},
	volume = {37},
	pages = {725}
}

@article{Kugel1975,
	title = {{Exchange interaction at triple orbital degeneracy}},
	author = {Kugel, K.~I. and Khomskii, D.~I.},
	journal = {Sov. Phys. Solid State},
	year = {1975},
	month = {Oct},
	volume = {17},
	pages = {285}
}

@book{Khomskii_book_2014,
	edition = {1},
	title = {{Transition Metal Compounds}},
	copyright = {https://www.cambridge.org/core/terms},
	isbn = {978-1-107-02017-7 978-1-139-09678-2},
	publisher = {Cambridge University Press},
	author = {Khomskii, Daniel I.},
	month = oct,
	year = {2014},
	doi = {10.1017/CBO9781139096782},
}

@article{Kugel1982,
	title = {{The Jahn-Teller effect and magnetism: transition metal compounds}},
	volume = {136},
	issn = {0042-1294, 1996-6652},
	url = {http://ufn.ru/ru/articles/1982/4/c/},
	doi = {10.3367/UFNr.0136.198204c.0621},
	number = {4},
	journal = {Usp. Fiz. Nauk},
	author = {Kugel, Kliment I. and Khomskii, Daniel I.},
	year = {1982},
	pages = {621},
}

@article{Tokura2000,
	title = {{Orbital Physics in Transition-Metal Oxides}},
	volume = {288},
	issn = {0036-8075, 1095-9203},
	url = {https://www.science.org/doi/10.1126/science.288.5465.462},
	doi = {10.1126/science.288.5465.462},
	number = {5465},
	journal = {Science},
	author = {Tokura, Y. and Nagaosa, N.},
	month = apr,
	year = {2000},
	pages = {462},
}

@article{Khaliullin2005,
	title = {{Orbital Order and Fluctuations in Mott Insulators}},
	volume = {160},
	issn = {0375-9687},
	url = {https://academic.oup.com/ptps/article-lookup/doi/10.1143/PTPS.160.155},
	doi = {10.1143/PTPS.160.155},
	journal = {Prog. Theor. Phys. Suppl.},
	author = {Khaliullin, Giniyat},
	year = {2005},
	pages = {155},
}

@article{Streltsov2017,
	title = {{Orbital physics in transition metal compounds: new trends}},
	volume = {60},
	issn = {1063-7869, 1468-4780},
	url = {https://ufn.ru/en/articles/2017/11/d/},
	doi = {10.3367/UFNe.2017.08.038196},
	number = {11},
	journal = {Phys.-Usp.},
	author = {Streltsov, S V and Khomskii, D I},
	month = nov,
	year = {2017},
	pages = {1121},
}

@article{Khomskii2022,
	title = {{Review—Orbital Physics: Glorious Past, Bright Future}},
	volume = {11},
	issn = {2162-8769, 2162-8777},
	url = {https://iopscience.iop.org/article/10.1149/2162-8777/ac6906},
	doi = {10.1149/2162-8777/ac6906},
	number = {5},
	journal = {ECS J. Solid State Sci. Technol.},
	author = {Khomskii, D. I.},
	month = may,
	year = {2022},
	pages = {054004},
}

@article{Radaelli2005,
	title = {{Orbital ordering in transition-metal spinels}},
	volume = {7},
	issn = {1367-2630},
	url = {https://iopscience.iop.org/article/10.1088/1367-2630/7/1/053},
	doi = {10.1088/1367-2630/7/1/053},
	journal = {New J. Phys.},
	author = {Radaelli, Paolo G},
	month = feb,
	year = {2005},
	pages = {53},
}

@article{Lee2010,
	title = {{Frustrated Magnetism and Cooperative Phase Transitions in Spinels}},
	volume = {79},
	issn = {0031-9015, 1347-4073},
	url = {https://journals.jps.jp/doi/10.1143/JPSJ.79.011004},
	doi = {10.1143/JPSJ.79.011004},
	number = {1},
	journal = {J. Phys. Soc. Jpn.},
	author = {Lee, Seung-Hun and Takagi, Hidenori and Louca, Despina and Matsuda, Masaaki and Ji, Sungdae and Ueda, Hiroaki and Ueda, Yutaka and Katsufuji, Takuro and Chung, Jae-Ho and Park, Sungil and Cheong, Sang-Wook and Broholm, Collin},
	month = jan,
	year = {2010},
	pages = {011004},
}

@incollection{Takagi2011,
	title = {{Highly Frustrated Magnetism in Spinels}},
	volume = {164},
	isbn = {978-3-642-10588-3 978-3-642-10589-0},
	url = {https://link.springer.com/10.1007/978-3-642-10589-0_7},
	booktitle = {{Introduction to Frustrated Magnetism}},
	publisher = {Springer Berlin Heidelberg},
	author = {Takagi, Hidenori and Niitaka, Seiji},
	editor = {Lacroix, Claudine and Mendels, Philippe and Mila, Frédéric},
	year = {2011},
	doi = {10.1007/978-3-642-10589-0_7},
	pages = {155},
}

@article{Tsurkan2021,
	title = {{On the complexity of spinels: Magnetic, electronic, and polar ground states}},
	volume = {926},
	issn = {03701573},
	shorttitle = {On the complexity of spinels},
	url = {https://linkinghub.elsevier.com/retrieve/pii/S0370157321001447},
	doi = {10.1016/j.physrep.2021.04.002},
	journal = {Phys. Rep.},
	author = {Tsurkan, Vladimir and Krug Von Nidda, Hans-Albrecht and Deisenhofer, Joachim and Lunkenheimer, Peter and Loidl, Alois},
	month = sep,
	year = {2021},
	pages = {1},
}

@article{Tsunetsugu2003,
	title = {{Magnetic transition and orbital degrees of freedom in vanadium spinels}},
	author = {Tsunetsugu, Hirokazu and Motome, Yukitoshi},
	journal = {Phys. Rev. B},
	volume = {68},
	issue = {6},
	pages = {060405},
	numpages = {4},
	year = {2003},
	month = {Aug},
	publisher = {American Physical Society},
	doi = {10.1103/PhysRevB.68.060405},
	url = {https://link.aps.org/doi/10.1103/PhysRevB.68.060405}
}

@article{Motome2004,
	title = {{Orbital and magnetic transitions in geometrically frustrated vanadium spinels: Monte Carlo study of an effective spin-orbital-lattice coupled model}},
	author = {Motome, Yukitoshi and Tsunetsugu, Hirokazu},
	journal = {Phys. Rev. B},
	volume = {70},
	issue = {18},
	pages = {184427},
	numpages = {22},
	year = {2004},
	month = {Nov},
	publisher = {American Physical Society},
	doi = {10.1103/PhysRevB.70.184427},
	url = {https://link.aps.org/doi/10.1103/PhysRevB.70.184427}
}

@article{Tchernyshyov2004,
	title = {{Structural, Orbital, and Magnetic Order in Vanadium Spinels}},
	author = {Tchernyshyov, O.},
	journal = {Phys. Rev. Lett.},
	volume = {93},
	issue = {15},
	pages = {157206},
	numpages = {4},
	year = {2004},
	month = {Oct},
	publisher = {American Physical Society},
	doi = {10.1103/PhysRevLett.93.157206},
	url = {https://link.aps.org/doi/10.1103/PhysRevLett.93.157206}
}

@article{Matteo2005,
	title = {{Orbital order in vanadium spinels}},
	author = {Di Matteo, S. and Jackeli, G. and Perkins, N. B.},
	journal = {Phys. Rev. B},
	volume = {72},
	issue = {2},
	pages = {020408},
	numpages = {4},
	year = {2005},
	month = {Jul},
	publisher = {American Physical Society},
	doi = {10.1103/PhysRevB.72.020408},
	url = {https://link.aps.org/doi/10.1103/PhysRevB.72.020408}
}

@article{Motome2005,
	title = {{Theory of Successive Transitions in Vanadium Spinels and Order of Orbitals and Spins}},
	volume = {160},
	issn = {0375-9687},
	url = {https://academic.oup.com/ptps/article-lookup/doi/10.1143/PTPS.160.203},
	doi = {10.1143/PTPS.160.203},
	journal = {Prog. Theor. Phys. Suppl.},
	author = {Motome, Yukitoshi and Tsunetsugu, Hirokazu},
	year = {2005},
	pages = {203},
}

@article{Sarkar2009,
	title = {{Proposed Orbital Ordering in ${\mathrm{MnV}}_{2}{\mathrm{O}}_{4}$ from First-Principles Calculations}},
	author = {Sarkar, S. and Maitra, T. and Valent\'{\i}, Roser and Saha-Dasgupta, T.},
	journal = {Phys. Rev. Lett.},
	volume = {102},
	issue = {21},
	pages = {216405},
	numpages = {4},
	year = {2009},
	month = {May},
	publisher = {American Physical Society},
	doi = {10.1103/PhysRevLett.102.216405},
	url = {https://link.aps.org/doi/10.1103/PhysRevLett.102.216405}
}

@article{Chern2010,
	title = {{Quantum $120\ifmmode^\circ\else\textdegree\fi{}$ model on pyrochlore lattice: Orbital ordering in ${\text{MnV}}_{2}{\text{O}}_{4}$}},
	author = {Chern, Gia-Wei and Perkins, Natalia and Hao, Zhihao},
	journal = {Phys. Rev. B},
	volume = {81},
	issue = {12},
	pages = {125127},
	numpages = {10},
	year = {2010},
	month = {Mar},
	publisher = {American Physical Society},
	doi = {10.1103/PhysRevB.81.125127},
	url = {https://link.aps.org/doi/10.1103/PhysRevB.81.125127}
}

@article{Adachi2005_MnV2O4,
	title = {{Magnetic-Field Switching of Crystal Structure in an Orbital-Spin-Coupled System: ${\mathrm{MnV}}_{2}{\mathrm{O}}_{4}$}},
	author = {Adachi, K. and Suzuki, T. and Kato, K. and Osaka, K. and Takata, M. and Katsufuji, T.},
	journal = {Phys. Rev. Lett.},
	volume = {95},
	issue = {19},
	pages = {197202},
	numpages = {4},
	year = {2005},
	month = {Nov},
	publisher = {American Physical Society},
	doi = {10.1103/PhysRevLett.95.197202},
	url = {https://link.aps.org/doi/10.1103/PhysRevLett.95.197202}
}

@article{Suzuki2007_MnV2O4,
	title = {{Orbital Ordering and Magnetic Field Effect in ${\mathrm{MnV}}_{2}{\mathrm{O}}_{4}$}},
	author = {Suzuki, T. and Katsumura, M. and Taniguchi, K. and Arima, T. and Katsufuji, T.},
	journal = {Phys. Rev. Lett.},
	volume = {98},
	issue = {12},
	pages = {127203},
	numpages = {4},
	year = {2007},
	month = {Mar},
	publisher = {American Physical Society},
	doi = {10.1103/PhysRevLett.98.127203},
	url = {https://link.aps.org/doi/10.1103/PhysRevLett.98.127203}
}

@article{Garlea2008_MnV2O4,
	title = {{Magnetic and Orbital Ordering in the Spinel ${\mathrm{MnV}}_{2}{\mathrm{O}}_{4}$}},
	author = {Garlea, V. O. and Jin, R. and Mandrus, D. and Roessli, B. and Huang, Q. and Miller, M. and Schultz, A. J. and Nagler, S. E.},
	journal = {Phys. Rev. Lett.},
	volume = {100},
	issue = {6},
	pages = {066404},
	numpages = {4},
	year = {2008},
	month = {Feb},
	publisher = {American Physical Society},
	doi = {10.1103/PhysRevLett.100.066404},
	url = {https://link.aps.org/doi/10.1103/PhysRevLett.100.066404}
}

@article{Nii2012_MnV2O4,
	title = {{Orbital structures in spinel vanadates $A$V${}_{2}$O${}_{4}$ ($A$ $=$ Fe, Mn)}},
	author = {Nii, Y. and Sagayama, H. and Arima, T. and Aoyagi, S. and Sakai, R. and Maki, S. and Nishibori, E. and Sawa, H. and Sugimoto, K. and Ohsumi, H. and Takata, M.},
	journal = {Phys. Rev. B},
	volume = {86},
	issue = {12},
	pages = {125142},
	numpages = {8},
	year = {2012},
	month = {Sep},
	publisher = {American Physical Society},
	doi = {10.1103/PhysRevB.86.125142},
	url = {https://link.aps.org/doi/10.1103/PhysRevB.86.125142}
}

@misc{Koyama2026,
	title={{Observation of complex orbital order in MnV$_2$O$_4$}}, 
	author={Chihaya Koyama and Shunsuke Kitou and Taishun Manjo and Yuiga Nakamura and Takeshi Hara and Naoyuki Katayama and Takuro Katsufuji and Ryotaro Arita and Yusuke Nomura and Hiroshi Sawa and Taka-hisa Arima},
	year={2026},
	eprint={2609.15060},
	archivePrefix={arXiv},
	primaryClass={cond-mat.str-el},
	url={https://arxiv.org/abs/2609.15060}, 
}

@article{Wheeler2010_MgV2O4,
	title = {{Spin and orbital order in the vanadium spinel ${\text{MgV}}_{2}{\text{O}}_{4}$}},
	author = {Wheeler, Elisa M. and Lake, Bella and Islam, A. T. M. Nazmul and Reehuis, Manfred and Steffens, Paul and Guidi, Tatiana and Hill, Adrian H.},
	journal = {Phys. Rev. B},
	volume = {82},
	issue = {14},
	pages = {140406},
	numpages = {4},
	year = {2010},
	month = {Oct},
	publisher = {American Physical Society},
	doi = {10.1103/PhysRevB.82.140406},
	url = {https://link.aps.org/doi/10.1103/PhysRevB.82.140406}
}

@article{Lee2004_ZnV2O4,
	title = {{Orbital and Spin Chains in ${\mathrm{Z}\mathrm{n}\mathrm{V}}_{2}{\mathrm{O}}_{4}$}},
	author = {Lee, S.-H. and Louca, D. and Ueda, H. and Park, S. and Sato, T. J. and Isobe, M. and Ueda, Y. and Rosenkranz, S. and Zschack, P. and \'I\~niguez, J. and Qiu, Y. and Osborn, R.},
	journal = {Phys. Rev. Lett.},
	volume = {93},
	issue = {15},
	pages = {156407},
	numpages = {4},
	year = {2004},
	month = {Oct},
	publisher = {American Physical Society},
	doi = {10.1103/PhysRevLett.93.156407},
	url = {https://link.aps.org/doi/10.1103/PhysRevLett.93.156407}
}

@article{Katsufuji2008_FeV2O4,
	title = {{Structural and Magnetic Properties of Spinel FeV$_2$O$_4$ with Two Ions Having Orbital Degrees of Freedom}},
	volume = {77},
	issn = {0031-9015, 1347-4073},
	url = {http://journals.jps.jp/doi/10.1143/JPSJ.77.053708},
	doi = {10.1143/JPSJ.77.053708},
	number = {5},
	journal = {J. Phys. Soc. Jpn.},
	author = {Katsufuji, Takuro and Suzuki, Takehito and Takei, Haruki and Shingu, Masao and Kato, Kenichi and Osaka, Keiichi and Takata, Masaki and Sagayama, Hajime and Arima, Taka-hisa},
	month = may,
	year = {2008},
	pages = {053708},
}

@article{Manjo2022,
	title = {{Do electron distributions with orbital degree of freedom exhibit anisotropy?}},
	volume = {3},
	issn = {2633-5409},
	url = {https://xlink.rsc.org/?DOI=D1MA01113H},
	doi = {10.1039/D1MA01113H},
	number = {7},
	journal = {Mater. Adv.},
	author = {Manjo, Taishun and Kitou, Shunsuke and Katayama, Naoyuki and Nakamura, Shin and Katsufuji, Takuro and Nii, Yoichi and Arima, Taka-hisa and Nasu, Joji and Hasegawa, Takumi and Sugimoto, Kunihisa and Ishikawa, Daisuke and Baron, Alfred Q. R. and Sawa, Hiroshi},
	year = {2022},
	pages = {3192},
}

@misc{Koyama_2026,
      title = {{Temperature evolution of orbital states with successive phase transitions in {FeV$_2$O$_4$}}},
      author = {Koyama, Chihaya and Nomura, Yusuke and Kitou, Shunsuke and Manjo, Taishun and Nakamura, Yuiga and Hara, Takeshi and Katayama, Naoyuki and Nii, Yoichi and Arita, Ryotaro and Sawa, Hiroshi and Arima, Taka-hisa},
      year={2026},
      eprint={2604.04398},
      archivePrefix={arXiv},
      primaryClass={cond-mat.str-el},
      url={https://arxiv.org/abs/2604.04398}
}

@article{Brink2001,
  title = {{Orbital ordering of complex orbitals in doped Mott insulators}},
  author = {van den Brink, Jeroen and Khomskii, Daniel},
  journal = {Phys. Rev. B},
  volume = {63},
  issue = {14},
  pages = {140416},
  numpages = {4},
  year = {2001},
  month = {Mar},
  publisher = {American Physical Society},
  doi = {10.1103/PhysRevB.63.140416},
  url = {https://link.aps.org/doi/10.1103/PhysRevB.63.140416}
}

@article{Mostovoy2002,
  title = {{Orbital Ordering in Frustrated Jahn-Teller Systems with 90\ifmmode^\circ\else\textdegree\fi{} Exchange}},
  author = {Mostovoy, M. V. and Khomskii, D. I.},
  journal = {Phys. Rev. Lett.},
  volume = {89},
  issue = {22},
  pages = {227203},
  numpages = {4},
  year = {2002},
  month = {Nov},
  publisher = {American Physical Society},
  doi = {10.1103/PhysRevLett.89.227203},
  url = {https://link.aps.org/doi/10.1103/PhysRevLett.89.227203}
}

@article{Mochizuki2004,
	title = {{Orbital physics in the perovskite Ti oxides}},
	volume = {6},
	issn = {1367-2630},
	url = {https://iopscience.iop.org/article/10.1088/1367-2630/6/1/154},
	doi = {10.1088/1367-2630/6/1/154},
	journal = {New J. Phys.},
	author = {Mochizuki, Masahito and Imada, Masatoshi},
	month = nov,
	year = {2004},
	pages = {154},
}

@article{Jackeli2008,
  title = {{Classical Dimers and Dimerized Superstructure in an Orbitally Degenerate Honeycomb Antiferromagnet}},
  author = {Jackeli, G. and Khomskii, D. I.},
  journal = {Phys. Rev. Lett.},
  volume = {100},
  issue = {14},
  pages = {147203},
  numpages = {4},
  year = {2008},
  month = {Apr},
  publisher = {American Physical Society},
  doi = {10.1103/PhysRevLett.100.147203},
  url = {https://link.aps.org/doi/10.1103/PhysRevLett.100.147203}
}

@article{Jackeli2009_Sr2VO4,
	title = {{Magnetically Hidden Order of Kramers Doublets in ${d}^{1}$ Systems: ${\mathrm{Sr}}_{2}{\mathrm{VO}}_{4}$}},
	author = {Jackeli, George and Khaliullin, Giniyat},
	journal = {Phys. Rev. Lett.},
	volume = {103},
	issue = {6},
	pages = {067205},
	numpages = {4},
	year = {2009},
	month = {Aug},
	publisher = {American Physical Society},
	doi = {10.1103/PhysRevLett.103.067205},
	url = {https://link.aps.org/doi/10.1103/PhysRevLett.103.067205}
}

@article{Feiner1997,
	title = {{Quantum Melting of Magnetic Order due to Orbital Fluctuations}},
	author = {Feiner, Louis Felix and Ole\ifmmode \acute{s}\else \'{s}\fi{}, Andrzej M. and Zaanen, Jan},
	journal = {Phys. Rev. Lett.},
	volume = {78},
	issue = {14},
	pages = {2799},
	numpages = {0},
	year = {1997},
	month = {Apr},
	publisher = {American Physical Society},
	doi = {10.1103/PhysRevLett.78.2799},
	url = {https://link.aps.org/doi/10.1103/PhysRevLett.78.2799}
}

@article{Kugel2015,
  title = {{Spin-orbital interaction for face-sharing octahedra: Realization of a highly symmetric SU(4) model}},
  author = {Kugel, K. I. and Khomskii, D. I. and Sboychakov, A. O. and Streltsov, S. V.},
  journal = {Phys. Rev. B},
  volume = {91},
  issue = {15},
  pages = {155125},
  numpages = {11},
  year = {2015},
  month = {Apr},
  publisher = {American Physical Society},
  doi = {10.1103/PhysRevB.91.155125},
  url = {https://link.aps.org/doi/10.1103/PhysRevB.91.155125}
}

@article{Koch-Janusz2015,
  title = {{Affleck-Kennedy-Lieb-Tasaki State on a Honeycomb Lattice from ${t}_{2g}$ Orbitals}},
  author = {Koch-Janusz, Maciej and Khomskii, D. I. and Sela, Eran},
  journal = {Phys. Rev. Lett.},
  volume = {114},
  issue = {24},
  pages = {247204},
  numpages = {5},
  year = {2015},
  month = {Jun},
  publisher = {American Physical Society},
  doi = {10.1103/PhysRevLett.114.247204},
  url = {https://link.aps.org/doi/10.1103/PhysRevLett.114.247204}
}

@article{Savary2021,
	title = {{Quantum loop states in spin-orbital models on the honeycomb lattice}},
	volume = {12},
	issn = {2041-1723},
	url = {https://www.nature.com/articles/s41467-021-23033-y},
	doi = {10.1038/s41467-021-23033-y},
	number = {1},
	journal = {Nat. Commun.},
	author = {Savary, Lucile},
	month = may,
	year = {2021},
	pages = {3004},
}

@article{Churchill2025,
	title = {{Microscopic roadmap to a Kitaev-Yao-Lee spin-orbital liquid}},
	volume = {10},
	issn = {2397-4648},
	url = {https://www.nature.com/articles/s41535-025-00744-9},
	doi = {10.1038/s41535-025-00744-9},
	number = {1},
	journal = {npj Quantum Mater.},
	author = {Churchill, Derek and Zhang, Emily Z. and Kee, Hae-Young},
	month = mar,
	year = {2025},
	pages = {26},
}

@article{Jackeli2009,
	title = {{Mott Insulators in the Strong Spin-Orbit Coupling Limit: From Heisenberg to a Quantum Compass and Kitaev Models}},
	author = {Jackeli, G. and Khaliullin, G.},
	journal = {Phys. Rev. Lett.},
	volume = {102},
	issue = {1},
	pages = {017205},
	numpages = {4},
	year = {2009},
	month = {Jan},
	publisher = {American Physical Society},
	doi = {10.1103/PhysRevLett.102.017205},
	url = {https://link.aps.org/doi/10.1103/PhysRevLett.102.017205}
}

@article{Ross2011,
	title = {{Quantum Excitations in Quantum Spin Ice}},
	author = {Ross, Kate A. and Savary, Lucile and Gaulin, Bruce D. and Balents, Leon},
	journal = {Phys. Rev. X},
	volume = {1},
	issue = {2},
	pages = {021002},
	numpages = {10},
	year = {2011},
	month = {Oct},
	publisher = {American Physical Society},
	doi = {10.1103/PhysRevX.1.021002},
	url = {https://link.aps.org/doi/10.1103/PhysRevX.1.021002}
}

@article{HanYan2017,
	title = {{Theory of multiple-phase competition in pyrochlore magnets with anisotropic exchange with application to ${\mathrm{Yb}}_{2}{\mathrm{Ti}}_{2}{\mathrm{O}}_{7}$, ${\mathrm{Er}}_{2}{\mathrm{Ti}}_{2}{\mathrm{O}}_{7}$, and ${\mathrm{Er}}_{2}{\mathrm{Sn}}_{2}{\mathrm{O}}_{7}$}},
	author = {Yan, Han and Benton, Owen and Jaubert, Ludovic and Shannon, Nic},
	journal = {Phys. Rev. B},
	volume = {95},
	issue = {9},
	pages = {094422},
	numpages = {39},
	year = {2017},
	month = {Mar},
	publisher = {American Physical Society},
	doi = {10.1103/PhysRevB.95.094422},
	url = {https://link.aps.org/doi/10.1103/PhysRevB.95.094422}
}

@article{Slater1954,
	title = {{Simplified LCAO Method for the Periodic Potential Problem}},
	author = {Slater, J. C. and Koster, G. F.},
	journal = {Phys. Rev.},
	volume = {94},
	issue = {6},
	pages = {1498--1524},
	numpages = {0},
	year = {1954},
	month = {Jun},
	publisher = {American Physical Society},
	doi = {10.1103/PhysRev.94.1498},
	url = {https://link.aps.org/doi/10.1103/PhysRev.94.1498}
}

@article{Lindgren1974,
	title = {The Rayleigh-Schrodinger perturbation and the linked-diagram theorem for a multi-configurational model space},
	volume = {7},
	issn = {0022-3700},
	url = {https://iopscience.iop.org/article/10.1088/0022-3700/7/18/010},
	doi = {10.1088/0022-3700/7/18/010},
	number = {18},
	journal = {J. Phys. B: At. Mol. Phys.},
	author = {Lindgren, I},
	month = dec,
	year = {1974},
	pages = {2441},
}

@article{Nemoto2000,
	title = {{Generalized coherent states for \textit{SU}(\textit{n}) systems}},
	volume = {33},
	issn = {0305-4470, 1361-6447},
	url = {https://iopscience.iop.org/article/10.1088/0305-4470/33/17/307},
	doi = {10.1088/0305-4470/33/17/307},
	number = {17},
	journal = {J. Phys. A: Math. Gen.},
	author = {Nemoto, Kae},
	month = may,
	year = {2000},
	pages = {3493},
}

@article{Kanamori1957,
	title = {{Theory of the Magnetic Properties of Ferrous and Cobaltous Oxides, I}},
	volume = {17},
	issn = {0033-068X},
	url = {https://academic.oup.com/ptp/article-lookup/doi/10.1143/PTP.17.177},
	doi = {10.1143/PTP.17.177},
	number = {2},
	journal = {Prog. Theor. Phys.},
	author = {Kanamori, Junjiro},
	month = feb,
	year = {1957},
	pages = {177},
}

@article{Kanamori1963,
	title = {{Electron Correlation and Ferromagnetism of Transition Metals}},
	volume = {30},
	issn = {0033-068X},
	url = {https://academic.oup.com/ptp/article-lookup/doi/10.1143/PTP.30.275},
	doi = {10.1143/PTP.30.275},
	number = {3},
	journal = {Prog. Theor. Phys.},
	author = {Kanamori, Junjiro},
	month = sep,
	year = {1963},
	pages = {275},
}

@article{Georges2013,
	title = {{Strong Correlations from Hund's Coupling}},
	volume = {4},
	issn = {1947-5454, 1947-5462},
	url = {https://www.annualreviews.org/doi/10.1146/annurev-conmatphys-020911-125045},
	doi = {10.1146/annurev-conmatphys-020911-125045},
	number = {1},
	journal = {Annu. Rev. Condens. Matter Phys.},
	author = {Georges, Antoine and Medici, Luca De' and Mravlje, Jernej},
	month = apr,
	year = {2013},
	pages = {137},
}

@article{Mizokawa1996,
	title = {{Electronic structure and orbital ordering in perovskite-type 3d transition-metal oxides studied by Hartree-Fock band-structure calculations}},
	author = {Mizokawa, T. and Fujimori, A.},
	journal = {Phys. Rev. B},
	volume = {54},
	issue = {8},
	pages = {5368},
	numpages = {0},
	year = {1996},
	month = {Aug},
	publisher = {American Physical Society},
	doi = {10.1103/PhysRevB.54.5368},
	url = {https://link.aps.org/doi/10.1103/PhysRevB.54.5368}
}

@article{Yang2020,
  title = {{${\mathrm{VI}}_{3}$: A two-dimensional Ising ferromagnet}},
  author = {Yang, Ke and Fan, Fengren and Wang, Hongbo and Khomskii, D. I. and Wu, Hua},
  journal = {Phys. Rev. B},
  volume = {101},
  issue = {10},
  pages = {100402},
  numpages = {5},
  year = {2020},
  month = {Mar},
  publisher = {American Physical Society},
  doi = {10.1103/PhysRevB.101.100402},
  url = {https://link.aps.org/doi/10.1103/PhysRevB.101.100402}
}

@article{Giannozzi2017,
	title = {{Advanced capabilities for materials modelling with Quantum ESPRESSO}},
	volume = {29},
	issn = {0953-8984, 1361-648X},
	url = {https://iopscience.iop.org/article/10.1088/1361-648X/aa8f79},
	doi = {10.1088/1361-648X/aa8f79},
	number = {46},
	journal = {J. Phys.: Condens. Matter},
	author = {Giannozzi, P and Andreussi, O and Brumme, T and Bunau, O and Buongiorno Nardelli, M and Calandra, M and Car, R and Cavazzoni, C and Ceresoli, D and Cococcioni, M and Colonna, N and Carnimeo, I and Dal Corso, A and De Gironcoli, S and Delugas, P and DiStasio, R A and Ferretti, A and Floris, A and Fratesi, G and Fugallo, G and Gebauer, R and Gerstmann, U and Giustino, F and Gorni, T and Jia, J and Kawamura, M and Ko, H-Y and Kokalj, A and Küçükbenli, E and Lazzeri, M and Marsili, M and Marzari, N and Mauri, F and Nguyen, N L and Nguyen, H-V and Otero-de-la-Roza, A and Paulatto, L and Poncé, S and Rocca, D and Sabatini, R and Santra, B and Schlipf, M and Seitsonen, A P and Smogunov, A and Timrov, I and Thonhauser, T and Umari, P and Vast, N and Wu, X and Baroni, S},
	month = nov,
	year = {2017},
	pages = {465901},
}

@article{Perdew1996,
	title = {Generalized Gradient Approximation Made Simple},
	author = {Perdew, John P. and Burke, Kieron and Ernzerhof, Matthias},
	journal = {Phys. Rev. Lett.},
	volume = {77},
	issue = {18},
	pages = {3865},
	numpages = {0},
	year = {1996},
	month = {Oct},
	publisher = {American Physical Society},
	doi = {10.1103/PhysRevLett.77.3865},
	url = {https://link.aps.org/doi/10.1103/PhysRevLett.77.3865}
}

@article{VanSetten2018,
	title = {{The PseudoDojo: Training and grading a 85 element optimized norm-conserving pseudopotential table}},
	volume = {226},
	issn = {00104655},
	url = {https://linkinghub.elsevier.com/retrieve/pii/S0010465518300250},
	doi = {10.1016/j.cpc.2018.01.012},
	journal = {Comput. Phys. Commun.},
	author = {Van Setten, M.J. and Giantomassi, M. and Bousquet, E. and Verstraete, M.J. and Hamann, D.R. and Gonze, X. and Rignanese, G.-M.},
	month = may,
	year = {2018},
	pages = {39},
}

@article{Nakamura2021,
	title = {{RESPACK: An ab initio tool for derivation of effective low-energy model of material}},
	volume = {261},
	issn = {00104655},
	url = {https://linkinghub.elsevier.com/retrieve/pii/S001046552030391X},
	doi = {10.1016/j.cpc.2020.107781},
	journal = {Comput. Phys. Commun.},
	author = {Nakamura, Kazuma and Yoshimoto, Yoshihide and Nomura, Yusuke and Tadano, Terumasa and Kawamura, Mitsuaki and Kosugi, Taichi and Yoshimi, Kazuyoshi and Misawa, Takahiro and Motoyama, Yuichi},
	month = apr,
	year = {2021},
	pages = {107781},
}

@article{Charlebois2021,
	title = {{Ab initio derivation of low-energy Hamiltonians for systems with strong spin-orbit interaction: Application to ${\mathrm{Ca}}_{5}{\mathrm{Ir}}_{3}{\mathrm{O}}_{12}$}},
	author = {Charlebois, Maxime and Mor\'ee, Jean-Baptiste and Nakamura, Kazuma and Nomura, Yusuke and Tadano, Terumasa and Yoshimoto, Yoshihide and Yamaji, Youhei and Hasegawa, Takumi and Matsuhira, Kazuyuki and Imada, Masatoshi},
	journal = {Phys. Rev. B},
	volume = {104},
	issue = {7},
	pages = {075153},
	numpages = {19},
	year = {2021},
	month = {Aug},
	publisher = {American Physical Society},
	doi = {10.1103/PhysRevB.104.075153},
	url = {https://link.aps.org/doi/10.1103/PhysRevB.104.075153}
}

@article{Pizzi2020,
	title = {{Wannier90 as a community code: new features and applications}},
	volume = {32},
	issn = {0953-8984, 1361-648X},
	shorttitle = {Wannier90 as a community code},
	url = {https://iopscience.iop.org/article/10.1088/1361-648X/ab51ff},
	doi = {10.1088/1361-648X/ab51ff},
	number = {16},
	journal = {J. Phys.: Condens. Matter},
	author = {Pizzi, Giovanni and Vitale, Valerio and Arita, Ryotaro and Blügel, Stefan and Freimuth, Frank and Géranton, Guillaume and Gibertini, Marco and Gresch, Dominik and Johnson, Charles and Koretsune, Takashi and Ibañez-Azpiroz, Julen and Lee, Hyungjun and Lihm, Jae-Mo and Marchand, Daniel and Marrazzo, Antimo and Mokrousov, Yuriy and Mustafa, Jamal I and Nohara, Yoshiro and Nomura, Yusuke and Paulatto, Lorenzo and Poncé, Samuel and Ponweiser, Thomas and Qiao, Junfeng and Thöle, Florian and Tsirkin, Stepan S and Wierzbowska, Małgorzata and Marzari, Nicola and Vanderbilt, David and Souza, Ivo and Mostofi, Arash A and Yates, Jonathan R},
	month = apr,
	year = {2020},
	pages = {165902},
}

@article{Joshi1999,
	title = {{Elementary excitations in magnetically ordered systems with orbital degeneracy}},
	author = {Joshi, A. and Ma, M. and Mila, F. and Shi, D. N. and Zhang, F. C.},
	journal = {Phys. Rev. B},
	volume = {60},
	issue = {9},
	pages = {6584--6587},
	numpages = {0},
	year = {1999},
	month = {Sep},
	publisher = {American Physical Society},
	doi = {10.1103/PhysRevB.60.6584},
	url = {https://link.aps.org/doi/10.1103/PhysRevB.60.6584}
}

@article{Lauchli2006,
	title = {{Quadrupolar Phases of the $S=1$ Bilinear-Biquadratic Heisenberg Model on the Triangular Lattice}},
	author = {L\"auchli, Andreas and Mila, Fr\'ed\'eric and Penc, Karlo},
	journal = {Phys. Rev. Lett.},
	volume = {97},
	issue = {8},
	pages = {087205},
	numpages = {4},
	year = {2006},
	month = {Aug},
	publisher = {American Physical Society},
	doi = {10.1103/PhysRevLett.97.087205},
	url = {https://link.aps.org/doi/10.1103/PhysRevLett.97.087205}
}

@article{GangChen2010,
	title = {{Exotic phases induced by strong spin-orbit coupling in ordered double perovskites}},
	author = {Chen, Gang and Pereira, Rodrigo and Balents, Leon},
	journal = {Phys. Rev. B},
	volume = {82},
	issue = {17},
	pages = {174440},
	numpages = {25},
	year = {2010},
	month = {Nov},
	publisher = {American Physical Society},
	doi = {10.1103/PhysRevB.82.174440},
	url = {https://link.aps.org/doi/10.1103/PhysRevB.82.174440}
}

\begin{acknowledgments}
We are grateful to C. Koyama, S. Kitou, T. Hara, T. Manjo, H. Sawa, and T.-H. Arima for fruitful discussions and for providing us with the crystal structure data.
This work was supported by JSPS KAKENHI (Grant Nos. JP23KJ0783, JP23H04869, JP25H01506, JP26K00646, JP26K21723), and JST-ASPIRE (Grant No. JPMJAP2512).
H.N. was supported by JSPS Overseas Research Fellowship.
\end{acknowledgments}

\section*{Author contributions}
The project was coordinated by Y.N. 
The calculation was performed by H.N. 
The manuscript was written by H.N. and Y.N. 

\section*{Competing interests}
The authors declare no competing interests.

\end{document}


\renewcommand{\thefigure}{S\arabic{figure}}
\renewcommand{\thetable}{S\arabic{table}}
\renewcommand{\thesection}{S\arabic{section}}
\renewcommand{\theequation}{S\arabic{equation}}

\begin{center}
	{\large \textmd{Supplementary Materials:} \\[0.3em]
		{\bfseries Spin-orbital exchange as a route to intertwined dipole-quadrupole orbital order in MnV$_2$O$_4$ under strong trigonal crystal field}}
\end{center}

\tableofcontents

\section{Multiorbital Hamiltonian}

In this section, we present the explicit forms of the multiorbital Hamiltonian introduced in the main text. 
We consider the electronic Hamiltonian for the $t_{2g}$ orbital manifold,
\begin{equation}\label{eq:app_H_full}
\mathcal{H} = \mathcal{H}_{\rm C} + \mathcal{H}_{\rm CEF}^{\rm tri} + \mathcal{H}_{\rm SOC} + \mathcal{H}_{\rm kin},
\end{equation}
where $\mathcal{H}_{\rm C}$ denotes the on-site Coulomb interaction, $\mathcal{H}_{\rm CEF}^{\rm tri}$ the trigonal crystal field, $\mathcal{H}_{\rm SOC}$ the atomic spin-orbit coupling (SOC), and $\mathcal{H}_{\rm kin}$ the electron hopping between neighboring V ions.

\subsection{Coulomb interaction}

The Coulomb interaction for the $t_{2g}$ orbitals is written in the Kanamori form~\cite{Kanamori1963, Georges2013},
\begin{equation}
\mathcal{H}_{\rm C} = \sum_i \left[ \frac{U-3J_{\rm H}}{2} N_i(N_i-1) -2J_{\rm H} \vb*{S}_i^2 -\frac{J_{\rm H}}{2} \vb*{L}_i^2 +\frac{5}{2} J_{\rm H} N_i \right],
\end{equation}
where $U$ and $J_{\rm H}$ denote the on-site Coulomb repulsion and Hund's coupling, and $N_i$, $\vb*{S}_i$, and $\vb*{L}_i$ denote the total electron number, total spin, and total orbital angular momentum operators at site $i$, respectively.

These operators are defined as
\begin{equation}
S_i^\mu = \vb*{c}_i^\dagger (l^0 \otimes \tfrac{1}{2}\sigma^\mu)\vb*{c}_i, \quad
L_i^\mu = \vb*{c}_i^\dagger (l^\mu \otimes \sigma^0)\vb*{c}_i, \quad
N_i = \sum_m n_{i,m},
\end{equation}
where
\begin{equation}
\vb*{c}_i = (\vb*{c}_{i,yz},\, \vb*{c}_{i,zx},\, \vb*{c}_{i,xy})^{\mathrm{T}}, \quad
\vb*{c}_{i,m} = (c_{i,m,\uparrow},\, c_{i,m,\downarrow}), \quad
n_{i,m} = \vb*{c}_{i,m}^\dagger \vb*{c}_{i,m}, 
\end{equation}
with $m = yz, zx, xy$ labeling the $t_{2g}$ orbitals.

The matrices $l^\mu$ represent the orbital angular momentum operators in the $t_{2g}$ basis $\{d_{yz}, d_{zx}, d_{xy}\}$,
\begin{equation}
l^x = \mqty( 0 & 0 & 0 \\ 0 & 0 & i \\ 0 & -i & 0 ), \quad
l^y = \mqty( 0 & 0 & -i \\ 0 & 0 & 0 \\ i & 0 & 0 ), \quad
l^z = \mqty( 0 & i & 0 \\ -i & 0 & 0 \\ 0 & 0 & 0 ).
\end{equation}

\subsection{Trigonal crystal field}

The trigonal crystal field is expressed as
\begin{equation}
\mathcal{H}_{\rm CEF}^{\rm tri} 
= \Delta_{\rm tri} \sum_i \vb*{c}_i^\dagger ((\hat{\vb*{n}}_i\cdot\vb*{l})^2 \otimes \sigma^0)\vb*{c}_i,
\end{equation}
where $\vb*{n}_i$ denotes the local trigonal axis at site $i$, which depends on the sublattice. 
For sublattice 1, it is given by $\hat{\vb*{n}}_1 = (1,1,1)/\sqrt{3}$, and the corresponding orbital matrix $(\hat{\vb*{n}}_1\cdot\vb*{l})^2$ in the $t_{2g}$ basis $\{d_{yz},d_{zx},d_{xy}\}$ is given by
\begin{equation}
(\hat{\vb*{n}}_1\cdot\vb*{l})^2 = -\frac{1}{3} \mqty(0 & 1 & 1 \\ 1 & 0 & 1 \\ 1 & 1 & 0). 
\end{equation}
The matrices for the other sublattices are obtained by symmetry operations.
Here, $\Delta_{\rm tri}$ represents the energy splitting between the $a_{1g}$ singlet and the $e_g^{\pi}$ doublet induced by the trigonal crystal field.

\subsection{Spin-orbit coupling}

The SOC term is given by
\begin{equation}
\mathcal{H}_{\rm SOC} = \zeta \sum_i \sum_\mu \vb*{c}_i^\dagger (l^\mu \otimes \tfrac{1}{2} \sigma^\mu) \vb*{c}_i,
\end{equation}
where $\zeta$ is the SOC constant.

\subsection{Hopping term}

The hopping Hamiltonian is written as
\begin{equation}
\mathcal{H}_{\rm kin} = \sum_{\langle i,j \rangle} \vb*{c}_i^\dagger (T_{ij}\otimes\sigma^0)\vb*{c}_j + \mathrm{h.c.},
\end{equation}
where $T_{ij}$ is the hopping matrix between sites $i$ and $j$. 
The explicit form of $T_{ij}$ is given in the main text.

\section{Hopping matrix in different orbital bases}

In this section, we provide details of the hopping processes in different orbital bases.
We first consider the real orbitals spanning the $e_g^\pi$ manifold and then rewrite the hopping matrix in the $l_{\rm eff}^Z$ basis. 

\subsection{Real-orbital basis}

We first introduce a real-orbital basis for the $e_g^\pi$ manifold.
For the bond connecting sublattices 1 and 4, the corresponding orbitals are given by
\begin{equation}
\ket*{\psi_{1\alpha}} = \frac{1}{\sqrt{6}} (\ket*{yz}+\ket*{zx}-2\ket*{xy}), \quad
\ket*{\psi_{1\beta}} = \frac{1}{\sqrt{2}} (\ket*{yz}-\ket*{zx}),
\end{equation}
on sublattice 1, and
\begin{equation}
\ket*{\psi_{4\alpha}} = \frac{1}{\sqrt{6}} (-\ket*{yz}-\ket*{zx}-2\ket*{xy}), \quad
\ket*{\psi_{4\beta}} = \frac{1}{\sqrt{2}} (-\ket*{yz}+\ket*{zx}),
\end{equation}
on sublattice 4.

In this basis, the hopping matrix in $e_g^\pi$ manifold takes the diagonal form
\begin{equation}
T_{14}^{e_g^\pi} = \mqty(
t_\alpha & 0\\
0 & t_\beta),
\end{equation}
where
\begin{equation}
t_\alpha = \frac{-t_1-t_2+2t_3+4t_4}{3}, \quad
t_\beta = -t_1+t_2.
\end{equation}
Using the hopping integrals obtained from the DFT calculations in Table~I of the main text, we obtain
\begin{equation}
t_\alpha=-214~\mathrm{meV}, \quad
t_\beta=-158~\mathrm{meV}.
\end{equation}

The microscopic origin of these hopping processes can be understood schematically in terms of direct $d$-$d$ and oxygen-mediated hopping.
To illustrate this, we consider the direct $d$-$d$ and oxygen-mediated contributions, denoted by $t_\nu^{\rm dir}$ and $t_\nu^{\rm oxy}$, respectively.
Using the Slater-Koster parameters~\cite{Slater1954}, the direct contributions are given by
\begin{equation}
t_\alpha^{\rm dir} = \frac{1}{2}t_{dd\sigma} -\frac{1}{3}t_{dd\pi} +\frac{1}{6}t_{dd\delta}, \quad
t_\beta^{\rm dir} = -t_{dd\delta}.
\end{equation}
Thus, within this description, $t_\alpha$ contains a substantial direct-hopping contribution, dominated by the $dd\sigma$-hopping, whereas the direct contribution to $t_\beta$ is much smaller.

The oxygen-mediated hopping is generally affected by the trigonal distortion of the oxygen octahedra, which is present even in the cubic phase.
For a qualitative understanding within the present description, we neglect the effect of this distortion and consider an ideal oxygen octahedron.
In this limit, the oxygen-mediated contributions are given by
\begin{equation}
t_\alpha^{\rm oxy} = \frac{1}{2}\frac{(t_{pd\pi})^2}{\Delta_p}, \quad
t_\beta^{\rm oxy} = -\frac{(t_{pd\pi})^2}{\Delta_p},
\end{equation}
where $\Delta_p$ denotes the charge-transfer energy between the $d$ orbitals of V ions and the $p$ orbitals of O ions.
Within this description, the oxygen-mediated contribution to $t_\alpha$ has the opposite sign to its dominant direct contribution and partially cancels it, whereas $t_\beta$ is predominantly associated with the oxygen-mediated process.
This provides a qualitative microscopic picture of why $t_\alpha$ and $t_\beta$ can acquire comparable magnitudes and the same sign, as found for the DFT-derived hopping parameters of MnV$_2$O$_4$.

\subsection{$l_{\rm eff}^Z$ basis}

We next rewrite the hopping matrix in the $l_{\rm eff}^Z$ basis $\{+1,0,-1\}$.
Here, $l_{\rm eff}$ denotes the effective orbital angular momentum of the $t_{2g}$ manifold, which satisfies $l_{\rm eff} = -l$.
On sublattice 1, the local states are given by
\begin{equation}
\begin{split}
\ket*{l_{\rm eff}^{Z_1} = 0} &= \frac{1}{\sqrt{3}} (\ket*{yz} + \ket*{zx} + \ket*{xy}), \\
\ket*{l_{\rm eff}^{Z_1} = \pm 1} &= \pm \frac{1}{\sqrt{3}} \left( \ket*{yz} + e^{\pm 2\pi i /3} \ket*{zx} + e^{\mp 2\pi i /3} \ket*{xy} \right).
\end{split}
\end{equation}
The overall phase convention is chosen for convenience.
On sublattice 4, the relative phases are modified due to the different orientation of the local trigonal axis, leading to
\begin{equation}
\begin{split}
\ket*{l_{\rm eff}^{Z_4} = 0} &= \frac{1}{\sqrt{3}} (-\ket*{yz} - \ket*{zx} + \ket*{xy}), \\
\ket*{l_{\rm eff}^{Z_4} = \pm 1} &= \pm \frac{1}{\sqrt{3}} \left( -\ket*{yz} - e^{\pm 2\pi i /3} \ket*{zx} + e^{\mp 2\pi i /3} \ket*{xy} \right).
\end{split}
\end{equation}

In this basis, the hopping matrix for the bond connecting sublattices 1 and 4 is then expressed as
\begin{equation}
\tilde{T}_{14} =
\mqty(
t_a & \omega t_c & \omega^2 t_d \\
\omega^2 t_c & t_b & -\omega t_c \\
\omega t_d & -\omega^2 t_c & t_a
),
\end{equation}
where $\omega=e^{2\pi i/3}$, and the four real integrals are expressed in terms of the hopping integrals $t_1, t_2, t_3$, and $t_4$ defined in Eq.~(2) of the main text as
\begin{equation}
\begin{split}
t_a &= \frac{-2t_1+t_2+t_3+2t_4}{3}, \qquad
t_b = \frac{-2t_1-2t_2+t_3-4t_4}{3}, \\
t_c &= \frac{t_1+t_2+t_3-t_4}{3}, \qquad
t_d = \frac{-t_1+2t_2-t_3-2t_4}{3}.
\end{split}
\end{equation}
Within the low-energy $l_{\rm eff}^Z=\pm1$ subspace, the diagonal and off-diagonal hopping amplitudes $t_a$ and $t_d$ can also be related to the real-orbital hopping amplitudes introduced in the preceding subsection as
\begin{equation}
t_a=\frac{t_\alpha+t_\beta}{2}, \quad
t_d=\frac{t_\beta-t_\alpha}{2}.
\end{equation}
These relations follow from
\begin{equation}
\ket*{l_{\rm eff}^{Z_i}=\pm1} = \frac{e^{\mp i\pi/6}}{\sqrt{2}} (i\ket*{\psi_{i\alpha}} \pm\ket*{\psi_{i\beta}}).
\end{equation}

\begin{figure}[t]
	\centering
	\includegraphics[width=6cm]{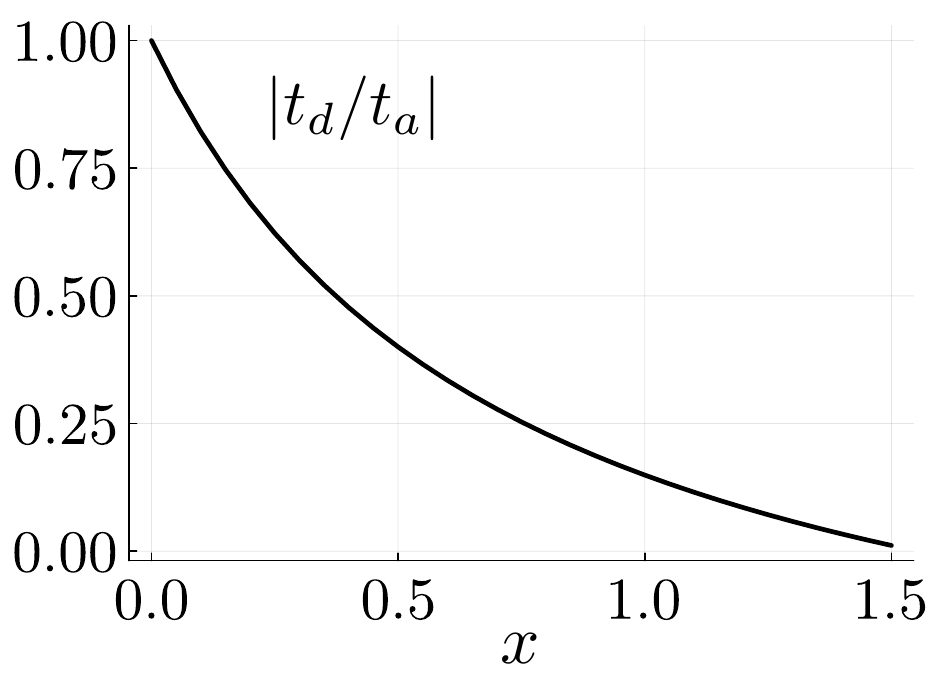}
	\caption{
		Ratio $|t_d/t_a|$ as a function of the hopping tuning parameter $x$. 
	}
	\label{fig:hopping}
\end{figure}

Using the hopping integrals obtained from the DFT calculations in Table~I in the main text, we obtain
\begin{equation}
t_a = -186~\mathrm{meV}, \quad
t_b = -34~\mathrm{meV}, \quad
t_c = -60~\mathrm{meV}, \quad
t_d = 28~\mathrm{meV}.
\end{equation}
These values show that $t_a$ is the dominant hopping integral in the $l_{\rm eff}^Z$ basis.
The relation to the real-orbital hopping amplitudes makes this hierarchy transparent: the comparable magnitudes and the same signs of $t_\alpha$ and $t_\beta$ lead to their addition in $t_a$ and cancellation in $t_d$.

For comparison, if only the dominant hopping $t_3$ is retained, one finds
\begin{equation}
t_a=t_b=t_c=-t_d=t_3/3,
\end{equation}
indicating the absence of a hierarchy among hopping processes.
By contrast, once the subdominant hoppings are included, the relative magnitude of $t_d$ is strongly suppressed compared with $t_a$. 
This behavior is shown in Fig.~\ref{fig:hopping}, where the ratio $|t_d/t_a|$ is plotted as a function of the tuning parameter $x$. 
Here, $x$ controls the relative strength of the subdominant hopping processes: the hopping matrix is decomposed as $T(x)=T'+xT''$, where $T'$ contains only the dominant hopping $t_3$, while $T''$ consists of the subdominant hoppings $t_1$, $t_2$, and $t_4$.

\section{Validation of the effective spin-orbital model}

In this section, we examine the validity of the effective spin-orbital model derived in the main text as a low-energy description of the multiorbital Hamiltonian.
We consider a two-site system and compare the results obtained by exact diagonalization of the full Hamiltonian [Eq.~(\ref{eq:app_H_full})] and the effective Hamiltonian $\mathcal{H}_{\rm eff}$ using the microscopic parameters for MnV$_2$O$_4$.
This provides a direct test of the strong-coupling expansion, with particular emphasis on the ground state relevant to the spin-orbital state discussed in the main text.

\begin{figure}[t]
	\centering
	\includegraphics[width=17cm]{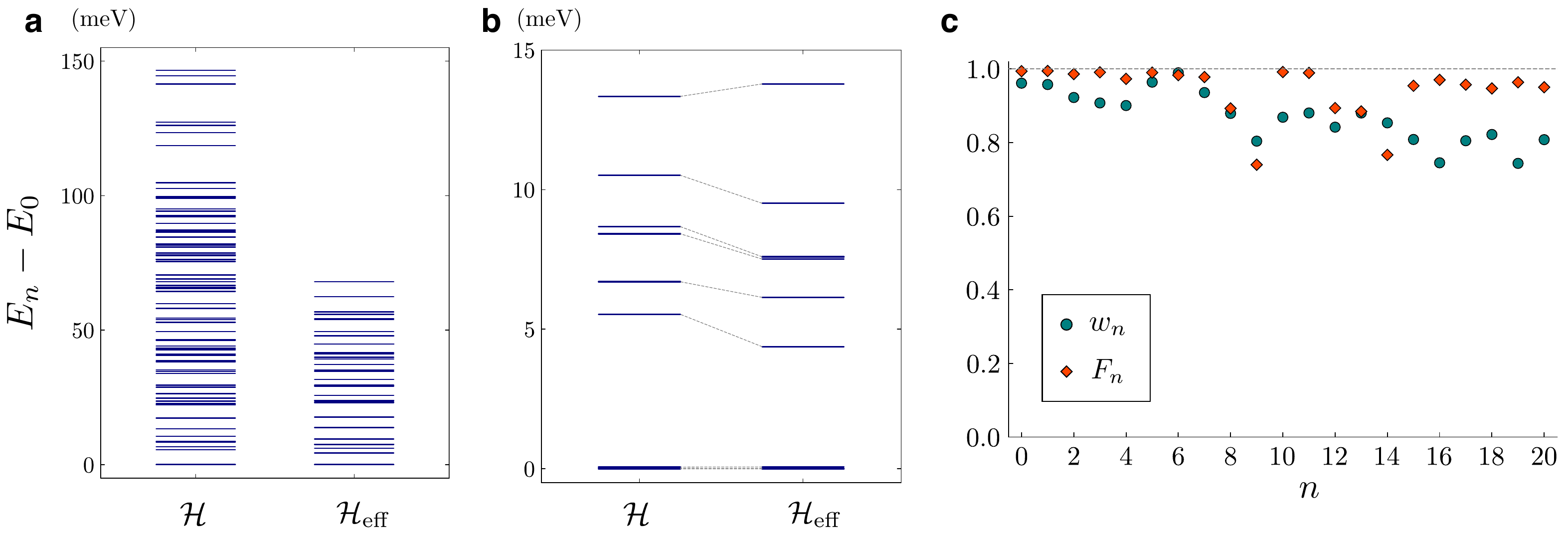}
	\caption{
		\textbf{a} Excitation spectra $E_n-E_0$ of the full Hamiltonian $\mathcal{H}$ and effective Hamiltonian $\mathcal{H}_{\rm eff}$ for a two-site system.
		For the full Hamiltonian, the lowest 81 states below the charge gap are shown, while all $6^2=36$ states are shown for the effective Hamiltonian.
		\textbf{b} Enlarged view of the low-energy spectra below $15$~meV, containing the lowest eight states ($n=0,\ldots,7$).
		\textbf{c} Projected weight $w_n$ and fidelity $F_n$ for the lowest 21 eigenstates ($n=0,\ldots,20$) of the full Hamiltonian.
		The microscopic parameters for MnV$_2$O$_4$ used in the main text are employed in all panels.
	}
	\label{fig:validation}
\end{figure}

Figure~\ref{fig:validation}(a) compares the excitation spectra of the full and effective Hamiltonians.
For the full Hamiltonian, we show the lowest 81 out of 495 states, which lie below the charge gap of approximately $1$~eV, while all $6^2=36$ states are shown for the effective model.
We note that, although the trigonal crystal-field splitting $\Delta_{\rm tri}=42$~meV sets the separation between the relevant multiplets in the single-site problem, no corresponding clear separation is observed in the two-site spectrum.

Figure~\ref{fig:validation}(b) shows an enlarged view of the spectra below $15$~meV, containing the lowest eight states ($n=0,\ldots,7$).
In this energy range, the full and effective Hamiltonians exhibit similar spectral structures, suggesting a correspondence between their low-energy states.
To characterize the correspondence between the two models, we define the projected weight
\begin{equation}
w_n = \ev*{\mathcal{P}}{\Psi_n^{\rm full}},
\end{equation}
where $\mathcal{P}$ projects onto the $6^2$-dimensional manifold with one $a_{1g}$ electron per site.
We also define the fidelity
\begin{equation}
F_n= \max_m \frac{|\mel*{\psi_m^{\rm eff}}{\mathcal{P}}{\Psi_n^{\rm full}}|^2}{w_n}. 
\end{equation}
where $\ket*{\psi_m^{\rm eff}}$ is an eigenstate of $\mathcal{H}_{\rm eff}$.
Figure~\ref{fig:validation}(c) shows $w_n$ and $F_n$ for the lowest 21 states of the full Hamiltonian.
The projected weight $w_n$ generally decreases with increasing excitation energy, indicating increasing mixing with states outside the manifold retained in the effective model.
Nevertheless, the lowest eight states all have $w_n>0.9$ and $F_n>0.97$, showing that the effective Hamiltonian provides an accurate description of these states and that higher-order corrections have only a minor effect on their wave functions.
These results validate the effective spin-orbital Hamiltonian as a low-energy description of the microscopic model for the realistic parameters of MnV$_2$O$_4$. 

Finally, we briefly examine the two lowest states in terms of the ``in''/``out'' description introduced in the main text.
Defining the corresponding local Ising states as
\begin{equation}
\ket*{\Uparrow} = \ket*{S^Z=+1,L_{\rm eff}^Z=+1}, \quad
\ket*{\Downarrow} = \ket*{S^Z=-1,L_{\rm eff}^Z=-1},
\end{equation}
and
\begin{equation}
\ket*{\psi_{\pm}} = \frac{1}{\sqrt{2}} \left( \ket*{\Uparrow,\Downarrow} \pm \ket*{\Downarrow,\Uparrow} \right), 
\end{equation}
we obtain
\begin{equation}
|\braket*{\psi_+}{\psi_0^{\rm eff}}|^2=0.803, \quad
|\braket*{\psi_-}{\psi_1^{\rm eff}}|^2=0.775.
\end{equation}
Including the spin canting found in the variational calculation by rotating the spin part of the local Ising states to the canting angle $\theta=79^\circ$ increases these overlaps to $0.841$ and $0.856$, respectively.
The energy splitting between these two states is $0.057$~meV for the full Hamiltonian and $0.062$~meV for the effective Hamiltonian, much smaller than the characteristic energy scales of the exchange interactions.
This small splitting indicates that the energy scale associated with intersite entanglement of these Ising variables is very small.
These results support the approximate site-product description adopted in the main text, in which the local Ising degrees of freedom are established primarily by the on-site SOC and their arrangement is determined by the exchange interactions.

\section{Order parameters and mean-field results}

\subsection{Definition and symmetry classification}

The ordered phases are characterized by three types of order parameters: 
(i) spin dipole moments, 
(ii) orbital dipole moments associated with $\tau^Z$, and 
(iii) orbital quadrupole moments associated with $\tau^X$ and $\tau^Y$.

Within the present mean-field framework, these order parameters are classified according to the irreducible representations of the $T_d$ point group~\cite{HanYan2017}. 
The spin order parameters decompose as $A_2 \oplus E \oplus 2T_1 \oplus T_2$, while the orbital dipole moments decompose into $A_2 \oplus T_1$ and the quadrupole moments into $E \oplus T_1 \oplus T_2$.
The explicit expressions are summarized in Tables~\ref{table:spin_order_parameters} and \ref{table:orbital_order_parameters}. 
For the orbital dipole moments, the $A_2$ representation corresponds to the all-in/all-out (AIAO) configuration, whereas the $T_1$ representation describes the two-in/two-out (2I2O) configuration. 
For the orbital quadrupole moments, the $E$ representation corresponds to a ferroquadrupolar order, while the $T_1$ and $T_2$ representations correspond to antiferroquadrupolar orders.

\subsection{Mean-field results}

\begin{figure}[t]
	\centering
	\includegraphics[width=10cm]{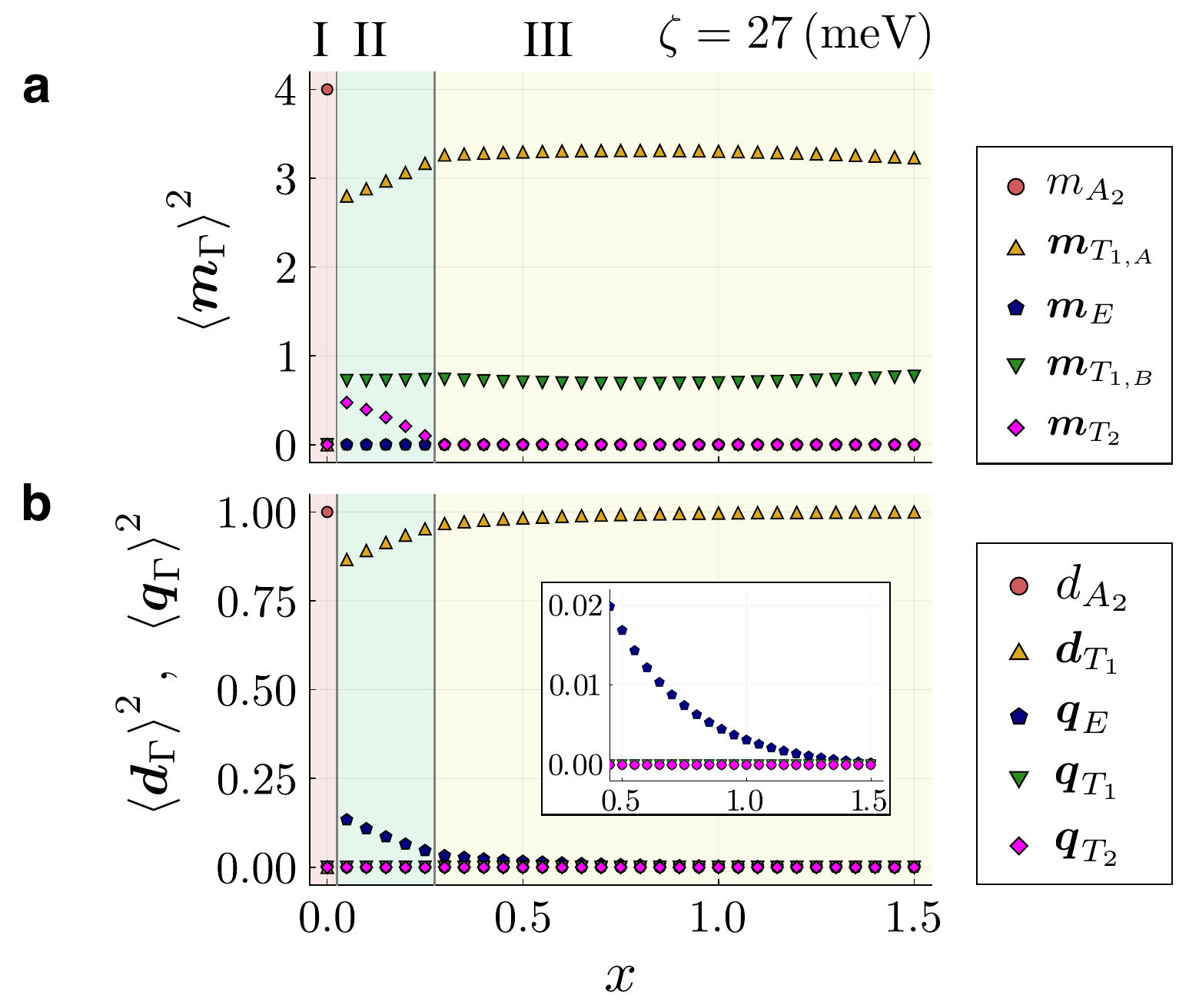}
	\caption{
		Squared amplitudes of the order parameters, $\ev*{\vb*{m}_\Gamma}^2$, $\ev*{\vb*{d}_\Gamma}^2$, and $\ev*{\vb*{q}_\Gamma}^2$, as a functions of the hopping tuning parameter $x$ at $\zeta = 27~\mathrm{meV}$. 
	}
	\label{fig:order_parameter}
\end{figure}

We summarize the mean-field results for the effective Hamiltonian. 
Three ordered phases are identified in the ground state. 
Phase~I is characterized by finite $m_{A_2}$ and $d_{A_2}$ without orbital quadrupole order, corresponding to the AIAO configuration for both spin and orbital moments. 
Phases~II and III exhibit 2I2O magnetic configurations with finite orbital dipole $\vb*{d}_{T_1}$ and quadrupole $\vb*{q}_E$ components. 
Phase~III includes the parameter set corresponding to MnV$_2$O$_4$ at $(x,\zeta)=(1,27~\mathrm{meV})$. 
The evolution of the order parameters as a function of the tuning parameter $x$ for $\zeta = 27~\mathrm{meV}$ is shown in Fig.~\ref{fig:order_parameter}. 

For comparison, a previous study~\cite{Chern2010} proposed an alternative state with $I4_1/a$ symmetry, characterized by $\vb*{m}_{T_1}$ and $\vb*{m}_E$ and accompanied by antiferroquadrupolar order $\vb*{q}_{T_1}$, which has been discussed in connection with the magnetic structure reported in experiments~\cite{Garlea2008_MnV2O4}. 
However, such a state is not obtained in the present calculations.

The magnetic structures are understood in terms of longitudinal and transverse components of the order parameters. 
The longitudinal components determine the AIAO or 2I2O configurations, while transverse components lead to spin canting and noncoplanar structures. 
Explicit expressions for representative configurations are provided below.

Representative spin configurations can be constructed as linear combinations of the symmetry components. 
For example, a canted 2I2O configuration is obtained from
\begin{equation}
\cos\alpha\, m^z_{T_{1,A}} + \sin\alpha\, m^z_{T_{1,B}}.
\end{equation}
The corresponding spin moments on the four sublattices are given by
\begin{equation}
\begin{split}
\mqty(S_1^x \\ S_1^y \\ S_1^z) 
&\propto 
\mqty(
\sqrt{2}\cos\alpha + \sin\alpha \\
\sqrt{2}\cos\alpha + \sin\alpha \\
\sqrt{2}\cos\alpha - 2\sin\alpha
), \quad 
\mqty(S_2^x \\ S_2^y \\ S_2^z) 
\propto 
\mqty(
-\sqrt{2}\cos\alpha - \sin\alpha \\
\sqrt{2}\cos\alpha + \sin\alpha \\
\sqrt{2}\cos\alpha - 2\sin\alpha
), \\[2mm]
\mqty(S_3^x \\ S_3^y \\ S_3^z) 
&\propto 
\mqty(
\sqrt{2}\cos\alpha + \sin\alpha \\
-\sqrt{2}\cos\alpha - \sin\alpha \\
\sqrt{2}\cos\alpha - 2\sin\alpha
), \quad
\mqty(S_4^x \\ S_4^y \\ S_4^z) 
\propto 
\mqty(
-\sqrt{2}\cos\alpha - \sin\alpha \\
-\sqrt{2}\cos\alpha - \sin\alpha \\
\sqrt{2}\cos\alpha - 2\sin\alpha
), 
\end{split}
\end{equation}
up to an overall normalization.
At $\alpha=-\arctan\sqrt{2}$, the in-plane components vanish, and the V spin configuration becomes collinear, with all V moments aligned along the global $z$ axis.

In Phase~II, an additional $m_{T_2}$ component can be included as
\begin{equation}
\cos\alpha\, m^z_{T_{1,A}} + \sin\alpha\cos\beta\, m^z_{T_{1,B}} + \sin\alpha\sin\beta\, m^z_{T_2},
\end{equation}
which further modifies the in-plane spin configuration.
For example, the spin at sublattice 1 is expressed in the global frame as
\begin{equation}
\mqty(S_1^x \\ S_1^y \\ S_1^z)
\propto
\mqty(
\sqrt{2}\cos\alpha + \sin\alpha(\cos\beta-\sqrt{3}\sin\beta) \\
\sqrt{2}\cos\alpha + \sin\alpha(\cos\beta+\sqrt{3}\sin\beta) \\
\sqrt{2}\cos\alpha - 2\sin\alpha\cos\beta
).
\end{equation}

\clearpage
\begin{table}[t]
	\caption{Spin order parameters classified by the irreducible representations of the point group $T_d$, expressed as linear combinations of spin operators in the local frame.}
	\vspace{5pt}
	\label{table:spin_order_parameters}
	\renewcommand{\arraystretch}{1.4}
	\setlength{\tabcolsep}{8pt}
	\centering
	\begin{tabular*}{\textwidth}{@{\extracolsep{\fill}} l c}
		\hline\hline
		& \\ [-5mm]
		\multicolumn{1}{l}{$m_{A_2}$} &
		$\dfrac{1}{2}(S_1^Z+S_2^Z+S_3^Z+S_4^Z)$ \\[2mm]
		\vspace{-15pt} &  \\ 
		\multicolumn{1}{l}{$\vb*{m}_E$} &
		$\dfrac{1}{2}\mqty(
		S_1^X+S_2^X+S_3^X+S_4^X \\
		S_1^Y+S_2^Y+S_3^Y+S_4^Y )$ \\[2mm]
		\vspace{-15pt} &  \\ 
		\multicolumn{1}{l}{$\vb*{m}_{T_{1,A}}$} &
		$\dfrac{1}{2}\mqty(
		S_1^Z+S_2^Z-S_3^Z-S_4^Z \\
		S_1^Z-S_2^Z+S_3^Z-S_4^Z \\
		S_1^Z-S_2^Z-S_3^Z+S_4^Z )$ \\[2mm]
		\vspace{-15pt} &  \\ 
		\multicolumn{1}{l}{$\vb*{m}_{T_{1,B}}$} &
		$\dfrac{1}{4}\mqty(
		2(S_1^X+S_2^X-S_3^X-S_4^X) \\
		-S_1^X+\sqrt{3}S_1^Y+S_2^X-\sqrt{3}S_2^Y-S_3^X+\sqrt{3}S_3^Y+S_4^X-\sqrt{3}S_4^Y \\
		-S_1^X-\sqrt{3}S_1^Y+S_2^X+\sqrt{3}S_2^Y+S_3^X+\sqrt{3}S_3^Y-S_4^X-\sqrt{3}S_4^Y )$ \\[2mm]
		\vspace{-15pt} &  \\ 
		\multicolumn{1}{l}{$\vb*{m}_{T_2}$} &
		$\dfrac{1}{4}\mqty(
		2(S_1^Y+S_2^Y-S_3^Y-S_4^Y) \\
		-\sqrt{3}S_1^X-S_1^Y+\sqrt{3}S_2^X+S_2^Y-\sqrt{3}S_3^X-S_3^Y+\sqrt{3}S_4^X+S_4^Y \\
		\sqrt{3}S_1^X-S_1^Y-\sqrt{3}S_2^X+S_2^Y-\sqrt{3}S_3^X+S_3^Y+\sqrt{3}S_4^X-S_4^Y )$ \\
		\vspace{-18pt} &  \\ 
		\hline\hline
	\end{tabular*}
\end{table}

\begin{table}[t]
	\caption{Orbital order parameters classified by the irreducible representations of the point group $T_d$, expressed as linear combinations of orbital pseudospin operators.}
	\vspace{5pt}
	\label{table:orbital_order_parameters}
	\renewcommand{\arraystretch}{1.4}
	\setlength{\tabcolsep}{8pt}
	\centering
	\begin{tabular*}{\textwidth}{@{\extracolsep{\fill}} l c}
		\hline\hline
		\multicolumn{2}{l}{(a) Dipolar order} \\ \hline
		\vspace{-15pt} &  \\ 
		\multicolumn{1}{l}{$d_{A_2}$} &
		$\dfrac{1}{2}(\tau_1^Z+\tau_2^Z+\tau_3^Z+\tau_4^Z)$ \\
		\vspace{-15pt} &  \\ 
		\multicolumn{1}{l}{$\vb*{d}_{T_1}$}
		& $\dfrac{1}{2}
		\mqty(\tau_1^Z +\tau_2^Z -\tau_3^Z -\tau_4^Z \\
		\tau_1^Z -\tau_2^Z +\tau_3^Z -\tau_4^Z\\
		\tau_1^Z -\tau_2^Z -\tau_3^Z +\tau_4^Z) $  \\
		\vspace{-18pt} &  \\ 
		\hline
		\multicolumn{2}{l}{(b) Quadrupolar order} \\ \hline
		\vspace{-15pt} &  \\ 
		\multicolumn{1}{l}{$\vb*{q}_{E}$} & 
		$\dfrac{1}{2}
		\mqty(\tau_1^X +\tau_2^X +\tau_3^X +\tau_4^X \\
		\tau_1^Y +\tau_2^Y +\tau_3^Y +\tau_4^Y)$ \\
		\vspace{-15pt} &  \\ 
		\multicolumn{1}{l}{$\vb*{q}_{T_1}$} & 
		$\dfrac{1}{4} 
		\mqty( 2(\tau_1^Y +\tau_2^Y -\tau_3^Y -\tau_4^Y) \\
		\sqrt{3} \tau_1^X - \tau_1^Y -\sqrt{3} \tau_2^X + \tau_2^Y +\sqrt{3} \tau_3^X -\tau_3^Y -\sqrt{3} \tau_4^X + \tau_4^Y \\
		-\sqrt{3}\tau_1^X - \tau_1^Y +\sqrt{3}\tau_2^X + \tau_2^Y +\sqrt{3} \tau_3^X +\tau_3^Y -\sqrt{3}\tau_4^X -\tau_4^Y )$  \\
		\vspace{-15pt} &  \\ 
		\multicolumn{1}{l}{$\vb*{q}_{T_2}$} & 
		$\dfrac{1}{4}
		\mqty( 2(\tau_1^X +\tau_2^X -\tau_3^X -\tau_4^X) \\
		-\tau_1^X -\sqrt{3}\tau_1^Y +\tau_2^X +\sqrt{3}\tau_2^Y -\tau_3^X -\sqrt{3}\tau_3^Y +\tau_4^X +\sqrt{3}\tau_4^Y \\
		-\tau_1^X +\sqrt{3}\tau_1^Y +\tau_2^X -\sqrt{3}\tau_2^Y +\tau_3^X -\sqrt{3}\tau_3^Y -\tau_4^X +\sqrt{3}\tau_4^Y )$ \\
		\vspace{-18pt} &  \\ 
		\hline\hline
	\end{tabular*}
\end{table}

\clearpage

\section{Dependence on Coulomb interaction parameters}

In the main text, we use $U=6$~eV and $J_{\rm H}=0.7$~eV~\cite{Mizokawa1996} to derive the effective spin-orbital Hamiltonian for MnV$_2$O$_4$.
Here, we examine the dependence of the exchange interactions and the resulting spin-orbital state on these Coulomb interaction parameters.
Since the on-site term $\mathcal{H}_{\lambda}$ is determined primarily by the SOC and trigonal crystal field, we focus on the variation of the exchange term $\mathcal{H}_{\rm ex}$ arising from changes in the parameters of $\mathcal{H}_{\rm C}$.

\subsection{Exchange interactions}

To examine the dependence on the Coulomb interaction parameters, we vary $U$ while keeping the ratio $J_{\rm H}/U$ fixed at its value used in the main text, and separately vary $J_{\rm H}/U$ at fixed $U=6$~eV.
All the other microscopic parameters are kept unchanged.

\begin{figure}[t]
	\centering
	\includegraphics[width=11cm]{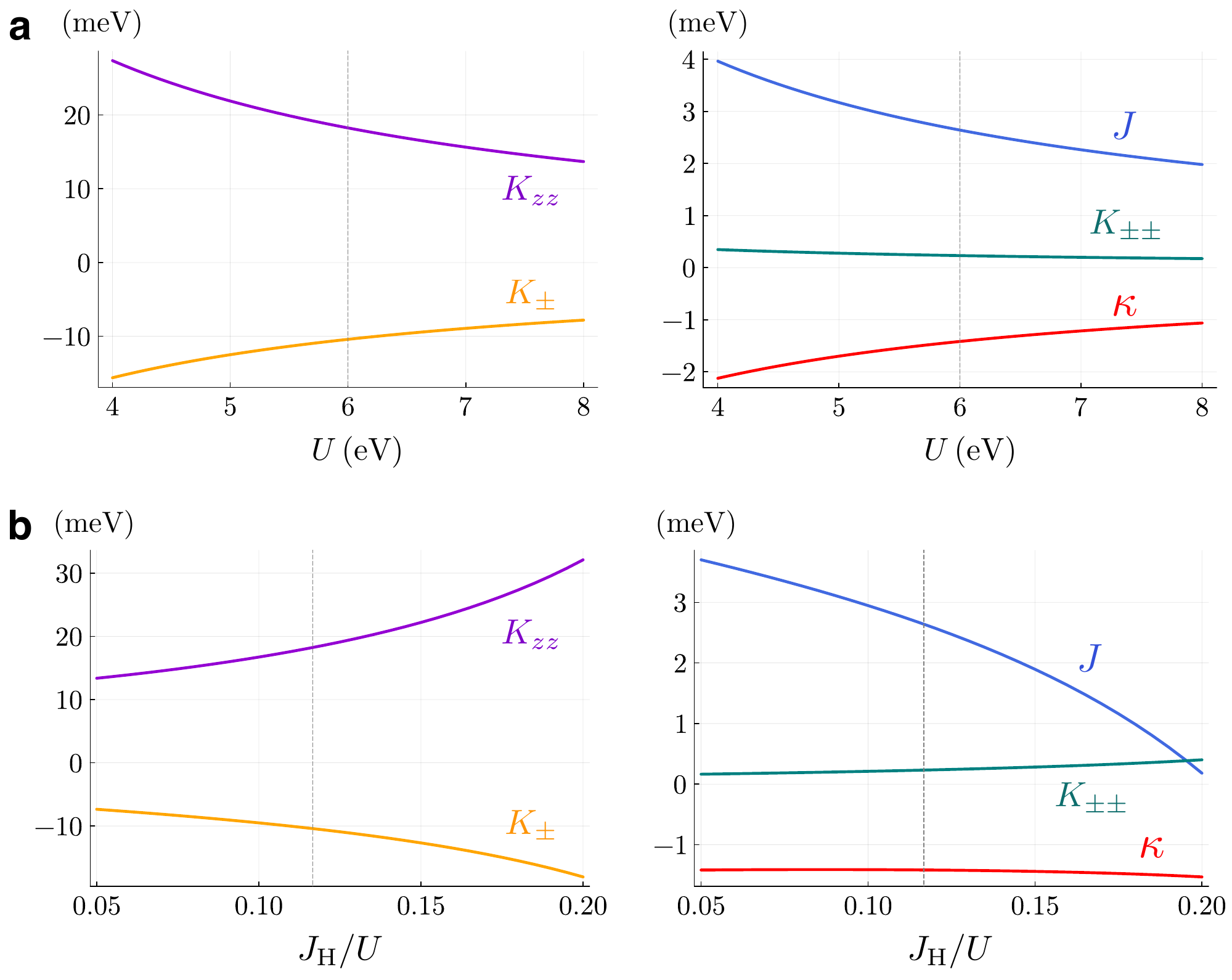}
	\caption{
		Dependence of the exchange parameters on the Coulomb interaction parameters.
		(a) Exchange parameters as functions of $U$ with $J_{\rm H}/U=0.7/6$ fixed.
		(b) Exchange parameters as functions of $J_{\rm H}/U$ at fixed $U=6$~eV.
		All the other microscopic parameters are fixed to the values used for MnV$_2$O$_4$ in the main text.
	}
	\label{fig:coulomb_exchange}
\end{figure}

Figure~\ref{fig:coulomb_exchange} shows the resulting effective exchange parameters as functions of $U$ and $J_{\rm H}/U$.
Although their magnitudes vary with the Coulomb interaction parameters, the relative hierarchy among the dominant exchange interactions remains largely unchanged over the parameter range considered.

\subsection{Spin-orbital state}

\begin{figure}[t]
	\centering
	\includegraphics[width=17cm]{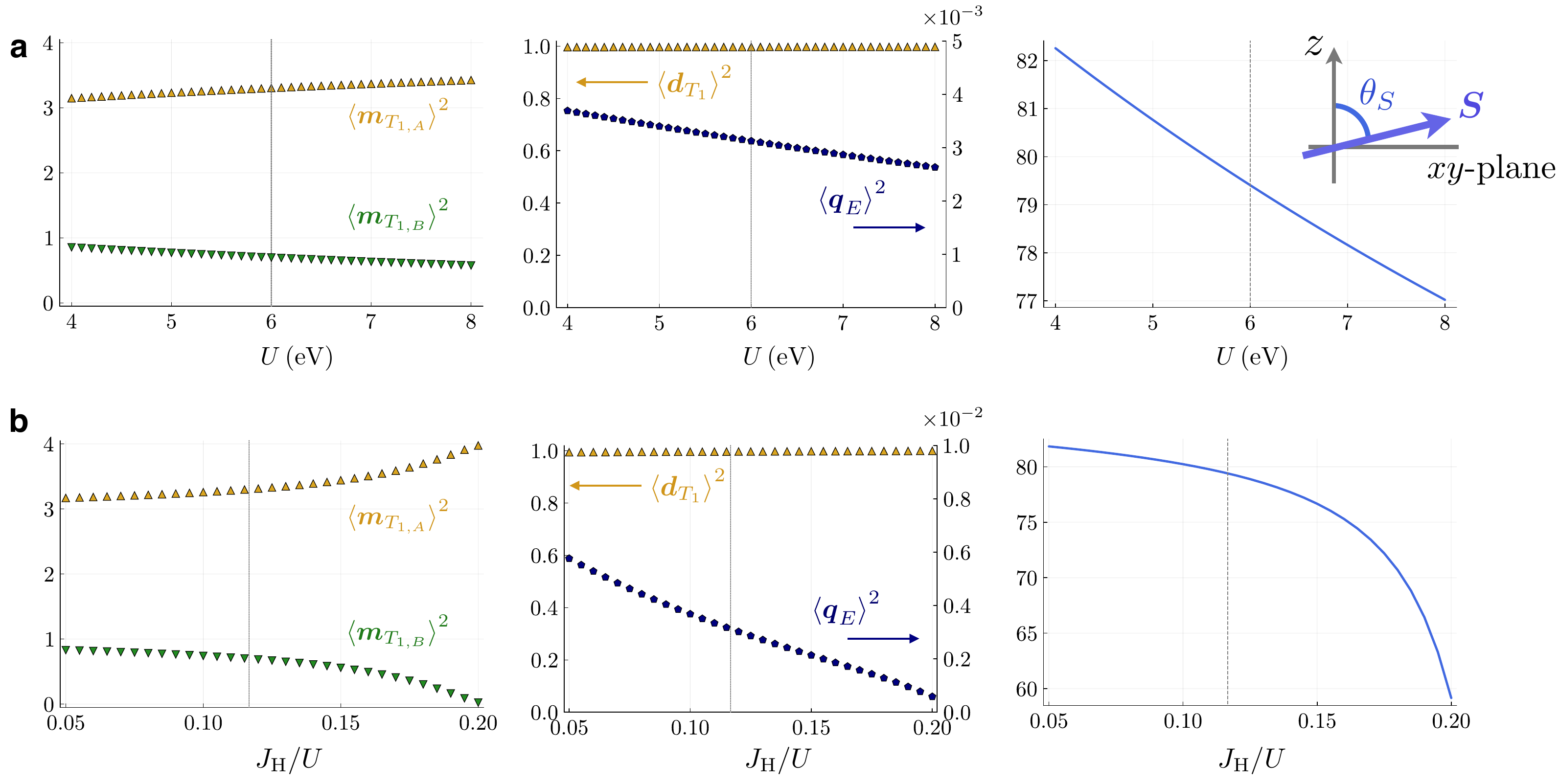}
	\caption{
		Dependence of the spin-orbital state for MnV$_2$O$_4$ on the Coulomb interaction parameters.
		(a) Results as functions of $U$ with $J_{\rm H}/U=0.7/6$ fixed.
		(b) Results as functions of $J_{\rm H}/U$ at fixed $U=6$~eV.
		From left to right, the panels show the spin order parameters, orbital order parameters, and spin canting angle $\theta_S$, respectively.
		The exchange parameters are evaluated for each value of $U$ or $J_{\rm H}/U$, while all other microscopic parameters are fixed to the values used in the main text.
	}
	\label{fig:coulomb_order}
\end{figure}

Using the exchange parameters obtained for each set of $(U,J_{\rm H})$, we perform the variational calculation at the parameter point corresponding to MnV$_2$O$_4$.
Figure~\ref{fig:coulomb_order} shows the resulting order parameters and spin canting angle as functions of $U$ and $J_{\rm H}/U$.
Throughout the parameter range considered, only $\vb*{m}_{T_1}$, $\vb*{d}_{T_1}$, and $\vb*{q}_E$ remain finite, while all the other order parameters remain zero, as in Phase~III discussed in the main text.
Thus, varying the Coulomb interaction parameters does not change the phase, although the magnitudes of the finite order parameters vary quantitatively.

In particular, the variations of $\vb*{m}_{T_{1,A}}$ and $\vb*{m}_{T_{1,B}}$ reflect the change in the spin canting angle.
The canting angle depends only weakly on $U$, whereas increasing $J_{\rm H}/U$ rotates the spins toward the global $z$ axis.
This behavior is consistent with the corresponding change in the Heisenberg exchange interaction shown in Fig.~\ref{fig:coulomb_exchange}: with increasing $J_{\rm H}/U$, the antiferromagnetic exchange becomes weaker.
Since the canting toward the global $xy$ plane is driven by the antiferromagnetic Heisenberg exchange, as discussed in the main text, its suppression reduces the canting away from the global $z$ axis.

These results demonstrate that the spin-orbital state obtained for MnV$_2$O$_4$ is robust against the variations of the Coulomb interaction parameters considered here.

\section{Linear flavor-wave analysis}

Here, we provide results of the linear flavor-wave (LFW) analysis described in the Methods.
We consider both the separable $\mathrm{SU}(3)\otimes\mathrm{SU}(2)$ local state used in the main text and the general SU(6) local state that allows on-site spin-orbital entanglement.
The calculations are performed for the DFT-derived model parameters.

\subsection{SU(3)$\otimes$SU(2) calculation}

We first perform the flavor-wave expansion around the $\mathrm{SU}(3)\otimes\mathrm{SU}(2)$ variational state used in the main text.
For a spin-$1$ degree of freedom, two local fluctuation modes are orthogonal to the SU(3) spin reference state, while one fluctuation mode is orthogonal to the SU(2) orbital reference state.
Thus, the $\mathrm{SU}(3)\otimes\mathrm{SU}(2)$ calculation contains three fluctuation modes per site.
The linear terms associated with these spin and orbital fluctuations vanish at the variational solution, confirming the stationarity of the reference state within the separable local-state space.

The six-dimensional local Hilbert space also contains two additional fluctuation modes that mix the spin and orbital sectors.
When these spin-orbital modes are included formally, their linear coefficients remain finite because the $\mathrm{SU}(3)\otimes\mathrm{SU}(2)$ variational state is not stationary against spin-orbital entangling variations.
We therefore retain only the three spin and orbital fluctuation modes in the $\mathrm{SU}(3)\otimes\mathrm{SU}(2)$ flavor-wave calculation.
For comparison, we also evaluated the quadratic Hamiltonian after formally including the two spin-orbital modes; the resulting ordered-moment corrections are reported below to illustrate the sensitivity of such an expansion to the nonstationary directions.

\subsection{SU(6) calculation}

To examine the effect of allowing on-site spin-orbital entanglement already at the mean-field (MF) level, we independently optimize a general SU(6) local state.
The resulting variational energy is lower than that of the separable $\mathrm{SU}(3)\otimes\mathrm{SU}(2)$ state by approximately $0.59~\mathrm{meV}$ per unit cell, while the qualitative character of the ordered state remains unchanged.
Around this SU(6) reference state, all five states orthogonal to the local reference state are retained as flavor-wave modes.
The linear coefficients vanish for all five modes, allowing a consistent quadratic expansion in the full local Hilbert space.

\subsection{Quantum correction to the ordered moments}

Table~\ref{tab:flavor_wave_moments} summarizes the quantum corrections to the spin and orbital ordered moments.
Because the four sublattices are symmetry related, we show the result for one representative sublattice.
We consider the magnitudes of the spin and orbital moments, $|\ev*{\vb*{S}}|$ and $|\ev*{\vb*{\tau}}|$, together with the quadruploar and dipolar components of the orbital moment, $|\vb*{\tau}_{\rm pl}|$ and $|\ev*{\tau^Z}|$, respectively.
For each quantity $O$, the relative quantum correction is defined as $(O_{\rm LFW}-O_{\rm MF})/O_{\rm MF}$, where $O_{\rm LFW}$ and $O_{\rm MF}$ denote the LFW and MF values, respectively.  

\begin{table}[t]
	\centering
	\caption{
		Ordered moments obtained at the mean-field level and after including linear flavor-wave corrections for one representative sublattice at the DFT-derived parameter set.
		The values in parentheses denote the relative changes from the corresponding mean-field values.
		The ``SU(3)$\otimes$SU(2)+SOF'' results are obtained by formally retaining the spin-orbital fluctuations (SOF) around the separable reference state and are shown only for comparison; because the linear terms in these modes do not vanish, this does not constitute a consistent quadratic expansion around the reference state.
	}
	\label{tab:flavor_wave_moments}
	\begin{tabular}{cccc}
		\hline\hline
		& SU(3)$\otimes$SU(2)
		& SU(3)$\otimes$SU(2)+SOF
		& SU(6) \\
		\hline
		$|\ev*{\vb*{S}}|$
		& $1.0000 \rightarrow 0.9232$ ($-7.68\%$)
		& $1.0000 \rightarrow 0.8890$ ($-11.0\%$)
		& $0.9968 \rightarrow 0.9116$ ($-8.55\%$)
		\\[1.5mm]
		$|\ev*{\vb*{\tau}}|$
		& $0.5000 \rightarrow 0.4588$ ($-8.24\%$)
		& $0.5000 \rightarrow 0.4163$ ($-16.75\%$)
		& $0.4970 \rightarrow 0.4145$ ($-16.60\%$)
		\\[1.5mm]
		$|\vb*{\tau}_{\rm pl}|$
		& $0.02804 \rightarrow 0.02573$ ($-8.24\%$)
		& $0.02804 \rightarrow 0.02657$ ($-5.23\%$)
		& $0.02078 \rightarrow 0.02025$ ($-2.53\%$)
		\\[1.5mm]
		$|\ev*{\tau^Z}|$
		& $0.4992 \rightarrow 0.4581$ ($-8.24\%$)
		& $0.4992 \rightarrow 0.4154$ ($-16.78\%$)
		& $0.4965 \rightarrow 0.4140$ ($-16.62\%$)
		\\
		\hline\hline
	\end{tabular}
\end{table}

Within the $\mathrm{SU}(3)\otimes\mathrm{SU}(2)$ calculation, the orbital moment is renormalized almost uniformly: the total, $XY$, and $Z$ components are all reduced by approximately $8.24\%$.
The spin moment is reduced by approximately $7.68\%$.
The SU(6) calculation gives a comparable reduction of the spin moment, approximately $8.55\%$.
The quantum correction affects the dipolar and quadrupolar orbital components differently: the dominant dipolar component is reduced by approximately $16.6\%$, whereas the smaller quadrupolar component is reduced by only approximately $2.5\%$.

Despite these quantitative differences, both flavor-wave calculations performed around variational states give moderate quantum corrections to the ordered moments and support the stability of the ordered state obtained in the mean-field analysis.
These moderate quantum corrections can be understood from the pronounced Ising-like character of the local spin-orbital states. 
The on-site SOC largely fixes the local spin-orbital configuration, suppressing quantum fluctuations away from the variational state.

\bibliographystyle{unsrt}
\bibliography{refs}